\documentclass[aps,psfig,preprint]{revtex4}
\usepackage{color}
\usepackage{graphics}
\usepackage{psfig}

\newcommand{\gr}{\nabla}

\newcommand{\Om}{\Omega}

\newcommand{\vph}{\varphi}

\newcommand{\fr}{\frac}

\newcommand{\pa}{\parallel}

\newcommand{\Pl}{\partial}

\newcommand{\ts}{\textstyle}

\newcommand{\bee}{\begin{equation}}
\newcommand{\ene}{\end{equation}}
\newcommand{\bem}{\begin{mathletters}}
\newcommand{\enm}{\end{mathletters}}
\newcommand{\bea}{\begin{array}}
\newcommand{\ena}{\end{array}}
\newcommand{\beea}{\begin{eqnarray}}
\newcommand{\enea}{\end{eqnarray}}

\newcommand{\bet}{\begin{tabbing}}
\newcommand{\ent}{\end{tabbing}}
\newcommand{\beb}{\begin{thebibliography}}
\newcommand{\enb}{\end{thebibliography}}

\newcommand{\lt}{\left}
\newcommand{\rt}{\right}

\newcommand{\fpar}[2]{\frac{{\ts \Pl \/ #1}}{{\ts \Pl \/ #2}}}

\newcommand{\npar}[3]{\frac{{\ts \Pl^{#1} \/ #2}}{{\ts \Pl \/ #3^{#1}}}}

\newcommand{\lsim}{\raisebox{-0.6ex}{\mbox{ $\stackrel{\ts <}{\ts \sim}$ }}}

\def\bfig{\begin{figure}}
\def\efig{\end{figure}}
\begin{document}
\DeclareGraphicsExtensions{.ps,.pdf,.png,.gif,.jpg}
\title{ Nonlinear saturation of magnetic curvature driven  Rayleigh Taylor instability in 
three dimensions}
 \author{$^a$Amita Das, $^a$Abhijit Sen,  $^a$Predhiman Kaw, \\
$^b$S. Benkadda and $^b$Peter Beyer }
\affiliation{$^a$Institute For Plasma Research, Bhat, Gandhinagar, 382428, India \\
$^b$Equipe Dynamique des Systemes Complexes, 13397 Marseille, France} 
\begin{abstract}
We present three dimensional fluid simulation results on the temporal evolution and 
nonlinear saturation of the magnetic curvature driven Rayleigh-Taylor (RT) instability. 
The model set of coupled nonlinear equations evolve the scalar electric field potential 
$\phi$, plasma density $n$ and the parallel component of the magnetic vector potential $\psi$. 
The simulations have been carried out in two limits, (i) a low resistivity case in which RT is the only linearly growing mode, and (ii) a high resistivity case where the drift wave is unstable and for which the  magnetic curvature parameter is set to  zero to ensure the absence of the RT growth. Our simulations show nonlinear stabilization in both these limits. The stabilization mechanism is similar to that observed in earlier two dimensional simulations, namely the generation of zonal shear flows which decorrelate the radially extended unstable modes. However the nature of the saturated nonlinear state in the 3d case differs from that of 2d in some important ways such as by having significant levels of power in short scales and by the presence of electromagnetic fluctuations. Though, in the linear regime the electromagnetic effects reduce the growth rates, in the nonlinear regime their presence  hinders the process of stabilization by inhibiting the process of zonal flow formation. Thus the parameter regime for which nonlinear stabilization takes place is considerably reduced in three dimensions.   
\end{abstract}
\maketitle
\section{Introduction}
The magnetic-curvature-driven Rayleigh-Taylor (MCD-RT) model is a useful paradigm for the study of 
long wave length nonlinear shear flow patterns such as zonal flows and streamer structures \cite{rtipr2}. These 
coherent potential patterns are believed to play an important role in determining the turbulent 
transport of matter and heat across field lines in magnetically confined plasma devices such as 
tokamaks, stellarators etc.  \cite{has,zlin,beer,dimits,syd,phd:iaea,rosen,busse,sagdeev,terry}. 
In the past few years a large number of 
investigations have been devoted to the elucidation 
of their characteristics and in understanding their contributions towards transport processes 
\cite{has,zlin,beer,dimits,syd,phd:iaea,rosen,busse,sagdeev,terry}. 
In the tokamak context, theoretical studies have mainly centered around the drift-wave model  
or its several variants such as the the ion temperature gradient (ITG) mode, the electron 
temperature gradient (ETG) mode \cite{dor1,dor2,singh},etc. The basic physics underlying the formation of these 
nonlinear structures is  the onset of a long scale modulational instability arising from 
the nonlinear parametric interaction of a large number of short scale fluctuations; this mechanism is in essence 
common to a number of earlier model calculations of strong plasma turbulence 
including that of the coupled interaction of Langmuir 
waves and ion-acoustic excitations \cite{zakh}. In a recent work \cite{rtipr2} the MCD-RT model was used to explore 
a very fundamental question pertaining to the formation of these nonlinear patterns, namely what 
determines the symmetry of the final
state - whether one gets zonal flows or streamer formations \cite{rtipr2}. The study highlighted the role of 
dissipative processes (namely the strength of the dissipation parameters - the flow viscosity $\mu$ and 
the diffusion coefficient $D$) in influencing the evolutionary path of the system towards a particular 
final symmetry state. Using extensive numerical simulation data it was possible to construct a consolidated 
``phase diagram'' in $D\;-\;\mu$ space which showed that low dissipation favoured the formation of zonal
 flows leading to saturated stationary states whereas high dissipation led to 
formation of radially elongated streamer flow patterns. The primary impact of the dissipation was on the 
initial evolution of the short scale fluctuations and the consequent distribution of their spectral 
power. High power in the short scales (due to weak dissipation) at the initial stages led to  stronger 
nonlinear generation of zonal flows whereas for high dissipation the streamers gained an upper hand and 
zonal flows were subdominant. We believe that such a unified and consolidated approach could be useful 
in gaining understanding of the evolution of similar nonlinear structures in tokamak transport models as 
well. Motivated by such considerations and also keeping in mind the fact that the RT model in a generic 
sense can provide valuable insights in a number of other experimental scenarios (e.g. currentless 
toroidal devices, ionospheric spread F irregularities etc.) we extend and further 
develop explorations on the nonlinear dynamics of this model. The earlier results reported in 
\cite{rtipr2,rtipr1}, were based on two dimensional simulations. In this paper we report a major extension of this 
model by making it fully three dimensional and also by including electromagnetic effects through the 
contributions of magnetic fluctuations. In terms of basic physics the extended model now introduces 
linear and nonlinear coupling to shear Alfven modes through finite $k_{\parallel}$ effects so that 
we now have a set of three coupled nonlinear equations that evolve the scalar electric field potential 
$\phi$, the plasma density $n$ and the parallel component of the magnetic vector potential $\psi$. We  
also have an additional dissipation parameter in the form of the resistivity coefficient $\eta_{s}$ 
which makes the drift wave branch linearly unstable in certain parameter ranges. We continue to 
explore the same fundamental issues in this generalized model, namely the existence of
saturated nonlinear states, their characteristics and the factors that influence their formation. 
Our approach is primarily numerical and we present extensive simulation results from our model equations 
to provide answers to the above issues. On the question of the existence of saturated states, past 
results from numerical explorations of the drift wave and allied models have been somewhat equivocal
and have indicated that electromagnetic effects tend to inhibit zonal flow development. In our numerical 
simulations of the fully electromagnetic RT model we find that saturated states still continue to,  exist although 
in a restricted parameter domain. Comparison with the earlier two dimensional results show a similarity 
in the saturation mechanism, namely through the excitation of zonal flows. However,  there are 
significant differences in some of the  characteristics of the saturated states.  Thus the three dimensional
 nonlinear states are found to possess a significantly higher  power level in short scales as compared 
to their 2d counterparts. Another significant difference is that the spatial structures of the 
potential and density fluctuations do  not  develop any functional correlations. In other words, 
unlike in the 2d case, the density evolution does not slave itself to the potential evolution. Our 
findings on the effect of electromagnetic fluctuations on RT turbulence are similar to those of drift 
wave models namely 
that they tend to inhibit the formation of zonal flows and thereby to considerably restrict the 
parametric domain of nonlinear stabilization. To highlight the role of the third dimension we have 
also explored a simplified limit of our model that can be termed as a three dimensional electrostatic 
model. In this simplified model the role of $k_{\parallel}$ and the resistivity factor $\eta_{s}$ are 
more transparent and we discuss their influence in the formation dynamics of the saturated states. 
In this limit we also demonstrate the stabilizing influence of the secondary Kelvin-Helmholtz 
instability \cite{chandra} in controlling the unlimited growth of streamers.\\

The paper is organized as follows. In the next section we present our generalized model equations and
and discuss its characteristics including its relation to the previous 2d model equations. We also
derive a simplified limit of two coupled equations representing the three dimensional electrostatic
model. Section III is devoted to delineating the properties of the linear modes of the model. This is 
done through numerical and approximate analytic solutions of the appropriate dispersion relations in 
various limits. This analysis also highlights the role of the various dissipation parameters and the 
parallel wavelength in the linear evolution stage of the system. We next present our nonlinear 
simulation results on the 3d electrostatic model in section IV and compare and contrast them with 
past 2d electrostatic results. The full electromagnetic simulation results of the generalized model 
are presented and discussed in section V. The paper ends with a summary of our main results and some 
concluding remarks in section VI.

\section{Model Equations}
The governing equations for the generalized model of the magnetic-curvature-driven Rayleigh-Taylor 
instability are derived along similar lines to that adopted for the previously investigated two 
dimensional set of equations \cite{rtipr2,rtipr1}. We use the fluid equations of continuity and momentum for 
the electrons and ions along with the quasi-neutrality condition viz. 
$\gr_{\perp} \cdot {\vec{J}}_{\perp} = - \gr_{\pa } J_{\pa}$. In addition we close the set by the Ohm's law 
which is the parallel component of the electron momentum equation without the electron inertia term. 
We use a slab representation in which the radial coordinate is represented by $x$, the poloidal by $y$ 
and the toroidal by $z$. The effect of curvature in equilibrium magnetic field is modeled by a $x$ 
dependent toroidal field  
$\vec{B}_{eq} = B_0(1-x/R)\hat{z}$ and the radial gradient of the equilibrium plasma 
density is represented by $n_0(x) = n_{00}exp(- x/L_n)$. We take the ions to be cold ($T_i =0$) 
but retain a finite electron temperature $T_e$. As a result of this the 
ion drift (flow) in equilibrium is zero but there is an equilibrium electron diamagnetic drift in the 
direction perpendicular to the magnetic field. Unlike  the 2d model we now retain electromagnetic 
perturbations ( $\vec{B}=\vec{B}_{eq}+\hat{z}\times \gr \psi$), where the perturbed magnetic field 
fluctuations are assumed to arise only from magnetic field line bending perturbations.  
Thus the vector potential has only a
$\hat{z}$ component i.e. $\vec{A} = -\psi \hat{z}$. The electric field is given by 
$\vec{E} = -\gr \phi + (1/c)\partial \psi/\partial t \hat{z}$. The scalar and vector potentials are 
finite only for the perturbations.  The total density is 
$N = n_0 + n_1$, thus $log N = -x/L_n +log (1+n_1/n_0) 
= -x/L_n + n$. The unit vector parallel to the magnetic field is given by 
$e_{\pa} =  \vec{B}/\mid B_0\mid 
= \hat{z} + (\hat{z} \times \gr \psi)/B_0 $. From standard perturbative expansions, the 
perpendicular component of the electron momentum equation yields the usual transverse flow 
velocities, namely the $\vec{E}\times \vec{B} $ drift  and the diamagnetic drift. The ion equation 
similarly leads to a $\vec{E}\times \vec{B} $ drift term as well as a polarization drift contribution in 
the perpendicular direction. These are substituted in the subsequent order equations to obtain the 
nonlinear evolution equations. We express the equations in a dimensionless form by normalizing the 
various physical quantities as follows. The  density is normalized by $n_{00}$, 
the electrostatic potential $\phi$ by $T_e/e$, time by $\Om_{i}$ and length by $\rho_s=c_s/\Om_i$. 
The vector potential $\psi$ is normalized by $B_0 \rho_s$. Our generalized set of model equations then 
consist of the following three coupled equations for the plasma density $n$, scalar electrostatic
 potential $\phi$ and the parallel (to the equilibrium magnetic field $\vec{B}_{eq}$) component of the 
vector potential $\psi$. 
\bee 
\fpar{n}{t} + V_g \fpar{n}{y} +
\lt(V_n - V_g \rt) \fpar{\vph}{y} +
\hat{z} {\bf \times} \vec{\gr} \vph \cdot \vec{\gr} n - 
V_A^2 \left\{\fpar{}{z} \gr_{\perp}^2 \psi + \hat{z} {\bf \times} \vec{\gr} \psi \cdot \vec{\gr} 
\gr_{\perp}^2 \psi \right\}= D \gr^2 n
\label{eq:1}
\ene
\bee 
\fpar{}{t} \gr^2 \vph + V_g \fpar{n}{y} +
\hat{z} {\bf \times} \vec{\gr}
\vph \cdot \vec{\gr} \gr^2 \vph  - 
V_A^2 \left\{\fpar{}{z} \gr_{\perp}^2 \psi + \hat{z} {\bf \times} \vec{\gr} \psi \cdot \vec{\gr} 
\gr_{\perp}^2 \psi \right\} = \mu \gr^4 \vph 
\label{eq:2}
\ene
\bee
\fpar{\psi}{t} + \fpar{}{z}(n - \vph) + V_n \fpar{\psi}{y} - \hat{z} {\bf \times} \vec{\gr}
\psi \cdot \vec{\gr} ( \vph - n)= \eta_s V_A^2 \gr_{\perp}^2 \psi 
\label{eq:3}
\ene
Here $V_g = c_s/(R \Om_i)$ is the gravitational drift arising
through the magnetic curvature terms, $V_n = c_s/(L_n \Om_i)$ is the
diamagnetic drift speed, $c_s$ is the ion acoustic speed, $R$ is
the major radius of curvature, $\Om_i$ is the ion cyclotron
frequency, $L_n$ is the equilibrium density scale-length and
$\mu$, and $D$ are the dynamical viscosity and  the  diffusion
coefficient respectively. 
Here  $V_A $ is the  Alfven velocity normalized to 
the sound velocity ($V_A^2 = v_a^2/c_s^2$). Thus the  plasma $\beta$ can be 
expressed as $\beta = 1/V_A^2$.
The coefficient of  resistivity defined as $\eta_s = \nu (c^2/\omega_p^2) 
(\omega_{ci}/v_a^2) = (\nu/\omega_p)(\omega_{ci}/\omega_p)(c^2/v_a^2) = \nu/\omega_{ce}$ is a 
dimensionaless parameter. 
Thus,  $\eta_s = 1.6 \times 10^{-13} n ln (\Lambda)/BT_e^{3/2}$, here $n$ is the plasma density, $B$ is 
the magnetic field 
in c.g.s system of units and $T_e$ is the electrom temperature in eV, $ln(\Lambda)$ is the Coulomb logarithm. 
The model set of Eqs.(\ref{eq:1}-\ref{eq:3}) has been derived earlier by Kaw (see \cite{shukla}). 
In the  present work we discuss in detail the linear and nonlinear features exhibited 
by these set of equations. 

	Comparing Eqs.(\ref{eq:1}-\ref{eq:2}) to our earlier 2d model equations, we see that the 
extended model has additional linear and nonlinear coupling  to the magnetic 
perturbation.  The coupling coefficient is proportional to $V_A^2$ 
(i.e. inversely proportional to $\beta$) as well as to the spatial variation in the parallel direction 
(i.e. to $k_{\parallel}$).
Equation (\ref{eq:3}) describes the time 
evolution of the magnetic fluctuation and is coupled both  to the density and 
potential fluctuations. In terms of basic physics the generalized model has an additional collective 
degree of freedom, namely  the shear Alfven modes and finite $k_{\parallel}$ effects bring about a 
linear and nonlinear coupling between them and the RT and drift modes. We also have an additional 
dissipation parameter in the system, namely  the resistivity coefficient $\eta_s$ appearing in the 
Ohm's law. We will discuss the linear properties of the model in greater detail in the next section. 

 Note that the two dimensional limit can be obtained by putting  $\partial /\partial z \rightarrow 0$ 
for which Eq.(\ref{eq:3}) gets 
totally decoupled from the equations for $n$ and $\phi$. In this limit,
it is easy to show from Eq.(\ref{eq:3}) that the magnetic energy ( $\sim \int{\mid \gr \psi^2}\mid d^3r$ ) simply decays
away at a rate proportional to $\eta_{s}V_A^2$. So in the two dimensional limit the magnetic energy has no 
role to play in the evolution of $n$ and $\phi$. Even when 3d effects are important the  electromagnetic effects 
can be negligible. This will happen when the evolution of $\psi$ becomes unimportant but the parallel current 
continues to remain  finite and provides  coupling to finite $k_{z}$ modes in the evolution equations for density 
and potential. Such a limit is possible when $\eta_s V_A^2 \gr^2 \psi >> \partial \psi/\partial t$ 
(or $\eta_s V_A^2 k_{\perp}^2 >> \gamma$, the growth rate ). The requisite limiting procedure thus consists of  
letting $\psi \rightarrow 0$ but letting the parallel current contribution (proportional to 
$\gr^2_{\perp} \psi$) on the RHS of eq.(\ref{eq:3}) remain finite. Thus from (\ref{eq:3}) we have,
\bee
\gr^2_{\perp} \psi = \fr{1}{\eta_s V_A^2}\frac{\partial}{\partial z}(n - \vph)
\ene
Substituting for $\gr^2_{\perp} \psi$ in (\ref{eq:1}) and (\ref{eq:2}) we obtain, 
\bee \fpar{n}{t} + V_g \fpar{n}{y} +
\lt(V_n - V_g \rt) \fpar{\vph}{y} +
\hat{z} {\bf \times} \vec{\gr} \vph \cdot \vec{\gr} n - 
\fr{1}{\eta_s} \left\{\npar{2}{}{z} (n - \vph)\right\} = D \gr^2 n
\label{eq_es:1}
\ene
\bee
\fpar{}{t} \gr^2 \vph + V_g \fpar{n}{y} +
\hat{z} {\bf \times} \vec{\gr}
\vph \cdot \vec{\gr} \gr^2 \vph  - 
\fr{1}{\eta_s} \left\{\npar{2}{}{z} (n - \vph)\right\}  = \mu \gr^4 \vph 
\label{eq_es:2}
\ene
We will refer to the above two coupled set of evolution equations (Eqs.(\ref{eq_es:1},\ref{eq_es:2})) 
 as the 3d electrostatic model. 
The  set of Eqs.(\ref{eq_es:1},\ref{eq_es:2}) are  considerably simplified  in comparison with 
 Eqs.(\ref{eq:1},\ref{eq:2},\ref{eq:3}).  They  describe the coupling between only two variables, viz. 
density and the scalar potential with no electromagnetic effects. 
However, the influence of the third dimension is still present through $\partial/\partial z$ for 
finite values of the parallel component of the wave vector  $k_{\pa}$. 
Physically, this simplified limit can be understood as follows.  The perpendicular variation 
of the Rayleigh Taylor mode  produces the charging of magnetic  
field lines via the polarization drift effect.  Finite spatial variation in the parallel direction 
implies that the magnetic field line is charged differently at different points thereby promoting 
the flow of a parallel current. The magnetic field associated with this 
current is responsible for the electromagnetic perturbations. However, if the resistivity 
of the plasma is high ($\eta_s V_A^2 k_{\perp}^2 >> \gamma $ the linear growth rate),  
such a parallel current gets  heavily damped, consequently the magnetic field 
as well as the creation of the rotational electric field $\partial \psi/\partial t$ is negligible  (and 
hence the limit $\psi \rightarrow 0$). The perturbations are therefore essentially 
electrostatic in nature in this limit and hence can be considered as an appropriate three dimensional 
extension of the earlier 2d electrostatic model. It provides a simple means of carrying out a direct 
comparison with the 2d results in the presence of finite $k_{\parallel}$ and finite resitivity effects. 

In  the absence of $V_g$, the gravitational 
drift, the 3D electrostatic model Eqs.(\ref{eq_es:1},\ref{eq_es:2}) reduces to the well known Hasegawa Wakatani 
model \cite{hw}, studied in great detail for the understanding of electrostatic 
low frequency plasma turbulence phenomena in three dimensions. Its 2D variants, obtained by replacing  
$z$ derivative by a single scalar number has also attracted considerable attention \cite{hw}. 


There is an interesting scaling property displayed by  both 3d electrostatic as 
well as the generalized 3d electromagnetic equations  which we now wish to highlight. 
The equations remain   invariant under  the following scaling transformations: 
\bee
z \rightarrow \bar{z}/a; \hspace{0.5cm} \eta_s = a^2 \bar{\eta_s}; 
\hspace{0.5cm} \psi = a \bar{\psi}; \hspace{0.5cm} V_A^2 = \bar{V_A^2}/a^2
\label{em_scale}
\ene
Here $a$ is a scalar scaling factor. 
These scalings help in establishing  equivalence amidst   a wide class of  phenomena for which 
the parameters $\beta$, $\eta_s$ and the typical length scales along the equilibrium magnetic field direction are 
related according to the  above mentioned scaling relations. 
Note that the transformation leaves $k_z V_A$, $\psi V_A$ and $\eta_s V_A^2$ invariant. 
It  also leaves the 
total energy (as well as each of the individual components of energy, namely pressure, kinetic and the 
magnetic energy) as invariant. Although the field $\psi$ gets scaled, yet the magnetic energy 
(normalized to plasma thermal energy viz. $\tilde{b}^2/8\pi n T$ )
which is $ \int (\gr \psi)^2 d^3r/(2 \beta \int d^3r) = V_A^2 \int (\gr \psi)^2 d^3r/(2 \int d^3r) $ 
in our normalizations,  remains invariant. 
The scaling relationship helps in carrying out simulation for a convenient choice of the set 
of parameters $\beta$, $\eta_s$ and $L_z$ (the box length along the $\hat{z}$ direction which defines 
the typical size of the excitation scales along $\hat{z}$), which can later be related to the 
realistic set of values using the scaling coefficient $a$. 

We will explore the nonlinear states of the 3D electrostatic  model set Eqs.(\ref{eq_es:1},\ref{eq_es:2})
as well as the full generalized electromagnetic set given by Eqs.(\ref{eq:1},\ref{eq:2},\ref{eq:3}) in sections 
IV and V after discussing their linear properties in the next section.
 
\section{Linear Analysis}
The coupled set of equations (\ref{eq:1},\ref{eq:2},\ref{eq:3}) can be linearized 
and fourier analyzed to obtain the following  dispersion relation 
\beea
&-&i k_{\perp}^2 \omega^3 + \left\{(D+\mu + \eta_s V_A^2)k_{\perp}^4 + ik_{\perp}^2 k_y(V_g +V_n) \right\} 
\omega^2 \nonumber \\
&+& \left\{i(D\eta_s V_A^2 +D\mu + \eta_s V_A^2 \mu)k_{\perp}^6 + i k_z^2 V_A^2 k_{\perp}^2 (1+k_{\perp}^2) \right\} 
\omega  \nonumber \\
&+ &\left\{-k_{\perp}^4 k_y  \left[ (\eta_s V_A^2 + \mu )V_g + (D+ \mu)V_n\right] + i k_y^2 
(V_g^2 - V_g V_n - k_{\perp}^2 V_gV_n) \right\} \omega \nonumber \\
&-& D \eta_s V_A^2 \mu k_{\perp}^8 - (D+ \mu k_{\perp}^2)k_z^2 V_A^2  k_{\perp}^4 
- i \mu k_{\perp}^6k_y (\eta_sV_A^2 V_g + DV_n) 
+i k_{\perp}^2 k_z^2 V_A^2k_y (V_g - V_n) \nonumber \\
&-& \eta_s V_A^2 k_{\perp}^2 k_y^2 V_g(V_g - V_n) + 
\mu k_{\perp}^4k_y^2V_n V_g -i k_y^3V_g^2V_n + i k_y^3 V_gV_n^2 = 0
\label{dispersion}
\enea
The above dispersion relation contains three basic modes, namely, the drift wave,  the Rayleigh-Taylor mode and the 
shear-Alfven wave. This can be seen quite easily by setting all the dissipative coefficients to be zero 
(i.e. $D = \mu = \eta_s = 0 $) and by rearranging Eq.(\ref{dispersion}) in the following form,
\bee
(\omega - k_y V_n) \left\{\omega^2 - k_yV_g \omega +\fr{k_y^2}{k_{\perp}^2}
V_g (V_n - V_g) \right\} = k_z^2 V_A^2(1+k_{\perp}^2) \left\{\omega + 
\fr{k_y(V_g - V_n)}{(1+k_{\perp}^2)} \right\}
\label{sim:disp}
\ene
For $V_g = V_n = 0$ (i.e. in the absence of magnetic curvature and density gradients) 
Eq.(\ref{sim:disp})  gives the kinetic Alfven wave dispersion  relation. 
\bee
\omega^2 =  k_z^2 V_A^2 (1+k_{\perp}^2)
\label{alfven}
\ene 
When  $V_n$ is finite, $V_g = 0 $  and  $V_A^2 \rightarrow \infty$ (a low $\beta $ plasma) we have upon 
dividing Eq.(\ref{sim:disp}) by $V_A^2$,  
\bee
\omega = \fr{k_y V_n}{(1+k_{\perp}^2)}
\label{drift}
\ene
which is the drift wave dispersion relation. For this case the electrons have a Boltzmann distribution, i.e. the wave
time scales are in the regime of $\omega/k_z v_{th,e} < 1$.  
The two dimensional electrostatic Rayleigh Taylor growth rate can be recovered 
by putting $k_z^2 V_A^2 \rightarrow  0$ in (\ref{sim:disp}). 
\bee
\left\{\omega^2 - k_yV_g \omega + \fr{k_y^2}{k_{\perp}^2}
V_g (V_n - V_g) \right\} = 0
\label{rt2d}
\ene
For this mode the electrons act like a two dimensional fluid,  under the  
condition of $\omega/k_z v_{th,e} >  1$. This is the only mode which is unstable (has a finite 
 growth rate) in the nondissipative limit. 
At a  finite value of $k_z V_A$  this unstable mode gets coupled to the stable Alfven branch. 
We show in Fig.1 the variation of the real and imaginary parts of $\omega$ as a function of 
$k_zV_A$. The value of $k_x = k_y = 0.1$ has been chosen for the plot.  The other parameters are 
$V_g = 0.036$, $V_n = 0.8$, 
 $D = \mu = \eta_s = 0$. 
For small values of $k_zV_A$ we observe that the three roots of the cubic equation are essentially 
obtained by putting the right hand side of Eq.(\ref{sim:disp}) to be zero. 
This gives rise to one real root $\omega = k_y V_n$ (arising due to the balance between the  parallel 
electric field and the equilibrium pressure variation along the bent magnetic field lines ), and 
two complex roots of the Rayleigh Taylor mode. For the parameters $k << 1 $ (scale lengths 
longer  than $\rho_s$) we have ($\omega = \omega_r + i \omega_i$)
$$
\omega_i = \pm \fr{1}{2}\left\{4 \fr{k_y^2}{k_{\perp}^2} V_g (V_n - V_g) - k_y^2 V_g^2 \right\}^{1/2}
\approx \pm \fr{k_y}{k_{\perp}} \sqrt{V_g(V_n - V_g)}
$$ 
$$
\omega_r = \fr{1}{2}k_y V_g
$$
Clearly, for a choice of $V_g << V_n$ the real part of the frequency is much smaller 
than the growth rate i.e. $\omega_r << \omega_i$ and is negligible as the plot of Fig.1  
in the regime of small $k_zV_A$ shows. 
As $k_zV_A$ increases the right hand side of Eq.(\ref{sim:disp}) cannot be ignored. 
Figure 1 shows that with increasing $k_zV_A$, $\mid \omega_i \mid$ decreases and goes to zero at $k_zV_A = k_{zc}V_A$.  
The point $k_z = k_{zc}$ is in fact  the point of exchange of instability, as 
$\omega_r \approx  0$ remains close to zero upto this point for the Rayleigh Taylor branch and becomes 
finite for values of $k_z V_A > k_{zc}V_A$. The expression for $k_{zc}$ can thus be 
determined from Eq.(\ref{sim:disp}) by substituting $\omega = 0$. This gives 
\bee
k_{zc} = k_y \sqrt{V_n V_g}/ V_A k_{\perp}
\label{kzlim}
\ene 
The critical wavenumber $k_{zc}$, beyond which the growth rate 
vanishes, thus increases with increasing values of $V_n$, $V_g$ and $k_y$ but decreases 
with increasing values of $V_A$ and $k_x$. For the values chosen for these parameters  in  Fig.1 
we have $k_{zc}V_A = 0.12$. 
The limit on $k_z$ for instability is in fact identical to the 
threshold condition on plasma $\beta$ encountered 
in the context of ideal Magnetohydrodynamic (MHD) ballooning modes in toroidal devices like tokamaks. 
In the ballooning mode case, one generally seeks the critical value of 
of plasma beta ( $\beta = \beta_c$), beyond which the instability sets in for a fixed value of the parallel 
wavenumber. 
 Here, on the other hand,  we have fixed the value of $\beta$ and are seeking the threshold 
condition on $k_z$ the parallel wavenumber below which the instability exists.  
 This has been done keeping in view the identification of the linearly 
unstable modes for the 
three dimensional simulations, where a range of $k_z$ are present and 
$\beta = 1/V_A^2$ would be taken  as a parameter. However, the expression for $\beta_c$ encountered in the context 
of ballooning modes can be recovered from Eq.(\ref{kzlim}) by substituting $k_{zc} = 1/qR$, $k_y \sim k_{\perp}$, 
$1/V_A^2 = \beta_c$, $V_nV_g =1/RL_n = 1/Ra $, (as the density gradient scale length $L_n$ can be taken typically to be 
of the order of minor radius $a$). These substitutions in Eq.(\ref{kzlim}) then lead to the well known expression 
for the critical value of plasma beta as $\beta_c = a/q^2 R = \epsilon/q^2$, where $\epsilon $ is the aspect ratio. 

In  the region $k_z < k_{zc}$ the curve  $\omega_i$ vs. $k_z$  in Fig.1 
typically seems to have an   elliptical shape. This can be understood as follows, we know that in this 
region $\omega_r \rightarrow 0$, hence $\omega^2 = -\omega_i^2$. Considering the
regions where $\omega_i < k_y V_n$ and using $V_g << V_n$ we write the dispersion relation as 
\bee
\omega_i^2 + k_z^2 V_A^2 - \fr{k_y^2}{k_{\perp}^2} V_g V_n  = 0; 
\hspace{0.5cm} i.e. \hspace{0.5cm}
\omega_i^2 + (k_z^2 - k_{zc}^2) V_A^2 = 0.  
\label{elliptic}
\ene
which is an equation of  an ellipse. Interestingly, the dispersion relation simplifies to 
similar elliptic form as of  Eq.(\ref{elliptic}) for $\omega_i > k_y V_n$, with the only modification 
that in this case the term $k_z^2 V_A^2$ is multiplied 
by the factor of $(1+k_{\perp}^2)$. 

There are certain other features  exhibited by the plot of  Fig.1. The only root 
which has the finite real part  at $k_z = 0$ is  $ \omega_r = k_y V_n$, and   arises from the 
decoupled $\psi$ equation in this limit;
 at higher $k_z$ it approaches the stable 
shear Alfven branch with $\omega_r = k_z V_A$. This mode remains stable throughout, 
the imaginary part 
of this particular mode remains $=0$ in the entire $k_z$ domain. The real part of the both  RT branches  are zero for 
$k_z < k_{zc}$; however as $k_z$ is increased beyond $k_{zc}$, one  amongst them asymptotes towards  the drift wave 
dispersion relation $\omega_r = k_yV_n/(1+k_{\perp}^2)$ and the other approaches the complementary 
branch of the shear Alfven mode i.e. $\omega = -k_zV_A$. 

 We now investigate the effect  of  dissipation  on the frequency as well as on the 
 growth rate of the three modes.  
In Fig.2  we have plotted the growth rate and the real frequency with $k_zV_A$  
when the dissipative coefficients $\mu$ (subplot (a) and  (b))  and $\eta_s$ (subplot(c) and  (d)) are separately 
taken to be finite. A comparison with the corresponding non dissipative case 
(all other parameters being identical)  of Fig.1, clearly shows that a finite value of  $\mu$ does not 
 alter the real frequency and the threshold condition on 
$k_zV_A$ significantly. It, however,  causes an  overall reduction in growth rate.   
 An entirely different and interesting trend is observed with respect to 
the  dissipative coefficient  $\eta_s$. As the value 
of $\eta_s$ is increased the real frequency instead of  suddenly acquiring  a finite value beyond $k_zV_A = k_{zc}V_A$,
gradually starts deviating from zero   even before $k_z = k_{zc}$ to finally 
asymptote towards the drift $k_y V_n/(1+k_{\perp}^2)$ 
and the shear Alfven branch $-k_z V_A$ at large enough $k_zV_A$.  
The growth rate too unlike the case of $\eta_s = 0$ continues  to remain finite even for $k_z > k_{zc}$ i.e. 
the unstable domain of the $k_z$ space gets widened. This happens essentially because of the presence of 
resistivity driven modes;  viz. the resistive 'g' and the resistive drift mode. 
The expenditure of energy in causing the field line bending along the parallel direction (for $\omega/ k_zv_{th e} > 1$ 
modes )
leads to the stabilization of the RT instability for finite $k_z$. The field lines bend due to the parallel 
component of the current. In the presence of resistivity the parallel currents gets damped. This leads to the 
recovery of instability in the resistive time scales, and is the physical basis of the excitation of the resistive 
interchange  mode. The resistive drift wave arises in the regime $\omega < k_z v_{th e}$ where the $n$, $\vph$
 relation wants to be Boltzmann like but acquired deviations because of finite resistivity effects; 
the phase difference between $n$ and $vph$ permits energy exchange between the waves and the fluid and leads to the 
resistive drift wave instability. 

In Fig.3 we depict the features of resistivity driven branch  in more detail.  
It  shows the plot of real (solid lines) and imaginary (dashed lines)  part of the 
frequency for this branch.  The ideal results $\eta_s = 0$ are plotted in subplot(a) ($V_g = 0.0364$) 
and subplot(c) ($V_g = 0.01$) to be  compared with the finite resistivity $\eta_sV_A^2  = 1.6$ plots of subplots (b) 
and (d) respectively. As mentioned earlier a finite growth rate   
beyond $k_z = k_{zc}$  is because of  the two resistivity driven modes, namely the 
resistive interchange mode  and 
the resistive drift wave. The region where  $\omega_i >> \omega_r$  has the characteristics of  the  resistive interchange 
mode; in the opposite limit  $\omega_r >> \omega_i $ it is  the resistive drift wave \cite{kadom}.  
 The figure also shows that 
as we reduce the value of $V_g$ the resistive drift regime gets broadened, ($\omega_r$ asymptotes to the drift wave 
frequency at  a lower $k_z$).

In Fig.4 we have isolated the  growth rate of the resistive drift mode by 
choosing   $V_g = 0$. One can see from the plots that the growth rate due to this mode vanishes for 
$k_z = 0$. An  analytical  expression for  the growth rate of this particular mode can be obtained perturbatively 
in the small $\eta_s$ limit as we show below.   
We look at the 
dispersion relation of Eq.(\ref{dispersion}) in the limit of finite $\eta_s$. 
We put  $D = \mu = 0$ for simplification. The dispersion relation in this case 
is similar to Eq.(\ref{sim:disp}) but with an additional 
$\eta_s$ dependent term, which is responsible for instability. 
\bee
(\omega - k_y V_n + i \eta_s k_{\perp}^2 V_A^2) \left\{\omega^2 - k_yV_g \omega +\fr{k_y^2}{k_{\perp}^2}
V_g (V_n - V_g) \right\} = k_z^2 V_A^2(1+k_{\perp}^2) \left\{\omega + 
\fr{k_y(V_g - V_n)}{(1+k_{\perp}^2)} \right\}
\label{sim:disp_eta}
\ene
The effect of $\eta_s$ on the drift wave can be seen by 
taking the limit $V_A^2 \rightarrow \infty$ and $V_g = 0$ in Eq.(\ref{sim:disp_eta}).
\bee
\omega - \fr{k_y V_n}{(1 + k_{\perp}^2)} = i \fr{\eta_s k_{\perp}^2}{k_z^2 (1+k_{\perp}^2)} \omega^2
\label{q_disp}
\ene
Considering the term on right hand side as a  small correction (possible when $\eta_s$ is small and $k_z$ is large) 
the dependence of $\eta_s$ on $\omega$ can be obtained iteratively as    
\bee
\omega = \fr{k_y V_n}{(1 + k_{\perp}^2)} + i \fr{\eta_s k_{\perp}^2}{k_z^2} \fr{k_y^2 V_n^2}{(1+k_{\perp}^2)^3}
\label{drift:resis}
\ene
which  shows the resistive destabilization of the drift wave branch. The expression obtained above 
shows that the growth rate reduces as $k_z$ is increased. The small $k_z$ limit can be captured by ignoring $\omega$ 
in the quadratic dispersion relation of Eq.(\ref{q_disp}) giving 
\bee
\omega = k_z \left(\fr{k_y V_n}{\eta_s k_{\perp}^2} \right)^{1/2} \fr{1+i}{\sqrt{2}}
\label{drift:resis1}
\ene
This shows that the growth rate increases with $k_z$ indicating $\gamma$ vs. $k_z$ must pass through a maximum, as 
seen in Fig.4. 

We next look at the linear properties of the simplified three dimensional electrostatic model equations 
(Eq.(\ref{eq_es:1},\ref{eq_es:2})). This model  
basically contains the Rayleigh Taylor mode and the drift wave 
mode,  but  the coupling to the shear-Alfven branch  is absent. The dispersion 
relation for this set is 
\bee
\omega^2 + \omega \left\{-k_y V_g + i\fr{k_z^2}{\eta_s}\fr{1+k^2}{k^2} \right\} + \fr{k_y^2}{k^2}V_g(V_n - V_g) 
-i \fr{k_y k_z^2}{\eta_sk^2}(V_n - V_g) = 0
\label{es_disp}
\ene
Clearly, one recovers the two dimensional RT growth rate expression in the $k_z^2=0$ or 
$\eta_s \rightarrow \infty$ limit. 
The first order perturbative corrections show damping due to $\eta_s$ and add a perturbative correction 
to the real frequency of this mode. The drift wave dispersion relation can also be recovered provided 
$k_z$ is finite. Taking  $V_g = 0$ and considering $\eta_s$ to be small 
 the dominant balance is between the two $k_z^2$ dependent 
terms in (\ref{es_disp}) from which we obtain the standard drift wave dispersion relation of 
$\omega = k_y V_n/(1+k^2)$. Retaining 
the first order correction due to the remaining $\eta_s$ dependent 
term we obtain the same expression for the resistive destabilization 
as that of  Eq.(\ref{drift:resis}). 

The role of $\eta_s$ 
on the drift wave branch for the general 
non perturbative case can be gleaned from the  plot of Fig.4 for both the electromagnetic and the simplified 
electrostatic models. The upper subplot  for $\eta_sV_A^2 = 1.6$ clearly shows that the 
two growth rates obtained from the two models differ at 
small $k_zV_A$ but agree well  at large values of $k_zV_A$.  
 At small $k_zV_A$ the growth rate obtained from the fully electromagnetic dispersion relation 
is found to be smaller than  the electrostatic case, which basically shows that the 
addition of electromagnetic effects  cause stabilization. 
The two growth rates, however, show a
similar trend in both cases, namely, they first  increase with $k_z$ and reach a 
maximum value and then fall off with $k_z$. The decreasing trend with $k_z$ is captured by the 
perturbative expression of Eq.(\ref{drift:resis}), which  is valid 
when $\eta_s/k_z^2$ is small.  Furthermore, the lower subplot of the figure 
also shows that for large values of 
$\eta_sV_A^2 $ ($ = 16$) the agreement between the two models is excellent  over  
the entire range of $k_zV_A$. This is a further evidence of the fact
that the approximations used in the derivation of the simplified   
electrostatic model are very accurate in the limit of large  resistivity. 
Though the other two modes are essentially damped by the finite value of $\eta_s$, 
the resistive destabilization of the  drift  mode  can enhance the growth rate of 
the coupled system. 

To summarize, linear analysis of the system shows that inclusion of the third dimension introduces additional 
unstable modes and the presence of electromagnetic effects brings about a coupling to shear Alfven modes. It is 
then of interest to understand the nonlinear evolution characteristics of these modes and their evolution into 
possible saturated nonlinear states. The simplified electrostatic model can be useful in  isolating the physics 
due to finite $k_z$ (parallel variation ) from the electromagnetic characteristics. In the next two sections we 
present the nonlinear evolution studies of these two models with the help of numerical simulations.\\

\section{Nonlinear simulation results for the 3d-electrostatic model} 
We begin by presenting and discussing the results of numerical simulation studies of the simplified 
3d-electrostatic model represented by Eqs.(\ref{eq_es:1},\ref{eq_es:2}). The equations are evolved with the 
help of a fully dealiased pseudospectral scheme  \cite{rtipr2,rtipr1}. Most of 
the studies have been carried out with a resolution of $64 \times 64 \times 64$ fourier modes in 
the three directions. Some test studies were also carried out with a lower resolution of 
 $64 \times 64 \times 16$ modes. 

We first investigate the question of the existence of saturated nonlinear states. 
In our earlier two dimensional studies 
it was shown that, even in the absence of any boundary or initial condition related anisotropy, 
there are two distinct symmetry states to which the system goes in the 
nonlinear state. The system ultimately forms  growing streamer structures which 
are radially (along $x$ in the slab description) elongated or to poloidally ($y$ direction)
symmetric saturated zonal patterns. For a fixed value of the driving parameters $V_n$ and $V_g$, the condensation 
to these symmetry states was governed by the value of dissipative coefficients $D$ and $\mu$. 
We now investigate the role of three dimensional perturbations on the development of these symmetry patterns. 
We choose the perpendicular box dimensions as $L_x = L_y = 20\pi \rho_s$; 
so that there is no boundary related anisotropy in the perpendicular plane. 
$L_z$ is chosen as $1.25 \times 10^5 \rho_s$; this is to concentrate on low frequency modes of interest 
which are extended along the field lines and have $k_z/k_{\perp} << 1$. 
We choose the 
 parametric regimes close to the 2d 
case in order to study the effects of parallel scales $k_z$ and the resistivity parameter 
$\eta_s$ on the formation of nonlinear states. 
We chose $D = \mu = 0.1$, 
$V_{n}=0.8 $, $V_{g} = 0.036$, a parametric region that corresponds to saturated zonal patterns for the two 
dimensional case  
 and  have  made several simulation runs
for different values of $\eta_s$ ($\lsim 10^{-5}$) and starting with small amplitude random 
initial perturbations in $n$ and $\vph$. 
The chosen  values of $V_n$ and $V_g$ also correspond to those of the currentless toroidal device 
BETA on which several experiments on the MCD-RT have been  carried out. For this machine the typical value of 
$\eta_s = \nu/\omega_{ce} \sim  10^{-5}$ and the parallel scale lengths are typically of the order of 
 $L_z \approx 125 Mts.$ ( $L_z = 2 \pi n R$ with the major radii  $R \approx 45 cm.$
  and $n \approx 30 - 50 $ is the observed toroidal winding number of the magnetic field lines in this machine).
Our results are applicable to other values of $\eta_s$ and $k_z$ through the scaling arguments described with 
Eq.(\ref{em_scale}) and may also be applied to the spread F - region of the ionosphere, ELM region of tokamaks 
etc. (for more details please see the last section). 

A typical 3d representation of the initial random potential $\vph$ structure is shown in the form of a slice plot in 
Fig.5. These slice plots basically show the pattern of a particular field variable with the help of 
color (in color plots) or through shading (in gray plots) on  
various two dimensional  slice planes  of  the three dimensional space. The appropriate  slicing  
helps in the   three dimensional visualization of the  field pattern. For instance in the plot of Fig.5 (and also 
in all the subsequent slice plots presented in this paper) we have chosen five different slices of 
the  three dimensional volume of $L_x \times L_y \times L_z$. The different slices are the $ x$ vs. $y$ plane 
at $z = 0$ and  $z = L_z/2$; $y$ vs. $z$ plane at $x = L_x/2$ and $L_x$; and $x$ vs. $z$ plane at $y = 0$. 

The initial density $n$ field has a similiar structure containing many short scales in a random configuration. 
For the small amplitudes chosen initially  
the evolution is primarily governed by the linear terms. Thus, during the initial linear phase  
modes having maximal growth rate acquire the largest amplitudes and dominate the spectrum.  
Since the maximally growing modes are the  RT modes that have large $k_y$ but small $k_x$, we can
expect to see, in the early linear stages of development, the appearance of structures which are 
elongated along the $x$ direction. We see clear evidence of such structures in the slice plot for the potential 
at $t=30$ in Fig.5. As the amplitudes grow, the modes  start interacting and one expects the power to get 
transferred to linearly stable modes as well. In the present case we see such a phenenomenon too and find 
the power in potential $\vph$ field nonlinearly cascading towards long scales. Such a cascade towards 
long scales  is an intrinsic  property of the polarization drift  nonlinearity that is present in 
the evolution equation of $\vph$. 

We have carried out simulations both for large and small resistivity parameters $\eta_s$, and observe distinct 
difference in the two regimes. For large values of $\eta_s$, viz. $10^{-5}$,   nonlinear saturated zonal symmetry 
patterns in potential $\vph$ field are seen to form (see Fig.5). However, when $\eta_s$ is small 
$ = 1.26 \times 10^{-6}$ there is no saturation and growing streamer patterns are observed (Fig.5). 
Thus in three dimensions, as  the available phase space of the modes  get enhanced with the addition of finite 
$k_z$ modes, the resistivity parameter $\eta_s$ along with $D$ and $\mu$ determine the symmetry pattern 
of the potential structure in the nonlinear stage. 
 Fig.6 shows the simulation cases in the parametric space of 
$D$ vs. $1/\eta_s$ for which saturated states were achieved (by circles) and those for which only growing 
streamer patterns were observed by $+$ (plus) signs.  
This trend is consistent with the fact that at large values of $\eta_s$ the contribution of the additional 
linear and nonlinear terms become small in comparison with other terms 
 and the set of equations tend to reduce  to the previous two dimensional equations. This has been 
quantitatively illustrated in Fig.7, which shows the plot of the ratio of growth rate $\gamma$ with 
$k_z^2/\eta_s$ as a function of $k_z$ for the two values of $\eta_s$ yielding saturated zonal (dots) and 
growing streamers (+ sign). The range of $k_z$ shows the permissible parallel wavenumbers of the simulation 
after aliasing. It is clear from the plot that when $\eta_s = 1.26 \times 10^{-6}$, there are parallel scales 
for which the ratio drops below unity (signifying the dominance of the extra three dimensional terms in the 
evolution equation and consequently the dynamics being altered significantly. On the other hand for large values 
of $\eta_s = 10^{-5}$ there are no parallel scales for which the additional terms dominate, the dynamics thus is close to 
the two dimensional scenario yielding saturated structures.

However, there are a few  interesting  differences in the composition of the final saturated state 
for the two and three dimensional cases even though the addtional terms are merely small perturbative corrections 
for such  numerical runs. 
 In the two dimensional simulations the density field was observed to get slaved to  $\vph$ 
and it too displayed the formation of long scale structures with two distinct symmetries. For the three 
dimensional runs however, we 
see, from the plot of Fig.8, that the density field continues to be dominated by power in the short scales. 
The scatter plots between the density and the potential fields (see Fig.9) 
also does not show any evidence of functional relationships developing  between  
density and potential fields. The vorticity $\gr^2 \vph$ too, unlike the previous case, 
has significant power in short scales (Fig.8) and does not form any functional relationship with the potential 
$\vph$ field (see Fig.9). 

The non slaving of the density field can be understood by realising that due to   
parallel variations,  additional modes (drift waves etc.) having very different linear mode relationship 
amidst the  two fields compared to the RT modes get excited (e.g. typically $n_k \sim \vph_k$ for drift waves, 
whereas for RT mode the density and potential fields are essentially out of phase ). This may hinder the 
slaving process of density to the potential. 
Moreover, two dimensional set of equations conserve the following non dissipative integral invariant 
\bee
\int \int \left\{ (\gr \vph)^2 - \fr{V_g}{(V_n - V_g)} n^2\right\}d^2 r = const
\label{2d_inv}
\ene 
which clearly shows an establishment of  integral relationship between the density and the potential fields. 
Such a integral relationship is consistent with the possibility of the $n$ field getting slaved to 
$\vph$. 
The incorporation of 3d effects, however, rules out any integral constraint on the two fields.  We have in this case  
instead:  
\bee
\fr{1}{2}\fpar{}{t}\int \int \int \left\{ (\gr \vph)^2 - \fr{V_g}{(V_n - V_g)} n^2\right\}d^3r = -
\fr{1}{\eta_s} \int \int \int \left\{ \left( \vph + \fr{V_g n}{(V_n - V_g)}  \right)\npar{2}{}{z}(n-\vph) 
  \right\}d^3r
\label{3d_inv}
\ene
The dissipation parameter $\eta_s$ has been retained,  as the 3d effects here arise solely from 
$\eta_s$ dependent term.
Clearly, since the two fields do not satisfy any integral relationship in the 3d case, it is not  easy 
for the density field to get slaved to $\vph$. It thus seems that the  non existence of any integral 
constraint and the absence of  any functional relationship between $n$ and $\vph$ 
along with the fact that the nonlinear evolution of the density field is governed directly by the 
convective nonlinearity  viz. $\hat{z} \times \gr \vph \cdot \gr n$  
which cascades power towards short scales,  leads to the predominance of power in short scale fluctuations in 
$n$. The polarization nonlinearity influences the $n$ evolution only indirectly through $\vph$. 

It should be noted that in the above runs with finite $\eta_s$, the linear phase has two unstable modes - the RT 
mode and the drift mode which is made unstable by the resistivity. In order to understand the role of the unstable 
drift wave it is possible to isolate its behaviour by artificially turning off the RT mode. We have carried out 
such an investigation by setting $V_g=0$ and looking
at the nonlinear evolution of the drift modes. The value of the resistivity parameter was chosen to be 
$\eta_s = 10^{-5}$. Figure 10 shows a comparison of the growth and evolution of the total energy, 
the zonal and streamer powers 
for  RT mode and the resistive drift wave. The slower linear 
rise can be understood from the lower growth rate of the resistivity driven drift wave in comparison to the 
growth rate of the RT mode. 
It is interesting to observe that in the final saturated regime of the resistive drift case, there is no dominance of 
power in the zonal mode as observed in the context of RT.  
This leads to a 
characteristic mixed flow pattern in which one cannot clearly distinguish between the zonal and streamer symmetries. 
Such a saturated state of the potential fluctuation at $t = 450$ is shown as a slice plot in Fig.11. 
It is also interesting to observe that for the resistive drift wave the energy level in finite $k_z$ modes 
is an order of magnitude higher than that in the $k_z = 0$ modes.

In all of the above simulations we have  avoided introducing any perpendicular anisotropy 
associated with boundary and initial conditions. We restricted ourselves to those simulations for which the    
 perpendicular aspect ratio of the simulation box  was unity i.e. $L_x = L_y$. 
The simulation volume $L_x \times L_y \times L_z$ basically represents a small region of the entire plasma. 
By simulating over a small region one hopes to identify and understand  the basic 
features of turbulent excitations. The simulation, however,  gets  constrained by the choice of box sizes in a few ways; 
the longest scale length along a particular direction is determined by the box length along that direction. 
In some  cases as we would show  below, a  natural process of power cascade towards  
a long scale  asymmetric mode can get inhibited by a certain  choice of the aspect ratio of the simulation box size.   
The unlimited growth of streamers observed by us earlier  for the choice of aspect ratio unity is 
one such example. It is well known that a shear flow excites Kelvin - Helmoltz (KH)instability; however, in our simulations 
carried out with with $L_x = L_y$ we observe no development of secondary KH instability which could prevent  the unlimited 
growth of   streamers  in the parameter 
domain indicated by the  $+$ sign of the plot in Fig.6. This happens because the essential condition for 
the excitation of KH instability on  
streamer shear flow (with shear scale length of $L_y$) can never be met within  the restriction of square box size. 
The KH instability can be excited only when the perturbation scales (in the orthogonal direction) are longer 
than the background shear scale length. With $L_x = L_y$ there can be  no such mode to support such a secondary 
destabilization process of streamers. By relaxing the constraint of $L_x = L_y$, and choosing instead $L_x > L_y$  
(we chose $L_x = 4L_y$) we observe that the unlimited and unphysical growth of streamers is prevented.  
In Fig.12 we show a comparison of the evolution of total energy $\int \int(n^2 + (\gr \vph)^2 ) dx dy$  for 
the two cases, viz. $L_x = L_y$ (solid lines) and $L_x = 4L_y$ (dotted lines). 
The other parameters for the two cases are identical. The plot clearly shows that the    
when the box dimensions 
are identical the energy grows indefinitely and there is no saturation. On the other hand when 
$L_x$ is chosen to be longer than $L_y$  the energy saturates. This has important implications on 
transport. It shows clearly that  simulations with those  parameters which require  $L_x > L_y$ for saturation;
 excites flows with  radial scale length longer than the ones for which  the instability saturates for $L_x = L_y$. 
Since the radial decorrelation step size is essentially determined by the radial scale length 
of the structures, it implies that the transport will be high for the parametric regime ($+$ sign 
of Fig.6) which do not saturate for a square box size (aspect ratio unity).

To summarize, in this section we have shown that similar to the 2d simulations the 3d electrostatic model 
too is capable of supporting 
nonlinear saturated states that are dominated by long scale zonal flow patterns. Here too,  
even in the absence of any boundary or initial condition related bias,  the 
nonlinear evolution  lead to the 
condensation towards either saturated zonal flow patterns  or growing 
streamer formations depending on the strength of the various 
dissipation parameters.  The model has an additional 
dissipation parameter in the resistivity coefficient which plays a special and distinctly different role 
than  the other two dissipation parameters in the selection of the final stage. 
The density field in the 3d saturated cases  show  distinct features of short scale dominance 
and non slaving to the $\vph$ field;  which is distinctly different from the  
2d results. Furthermore, the resistivity parameter 
also leads to the excitation of additional instabilities e.g. resistive interchange and the resistive drift waves. 
Simulation studies on the unstable resistive drift wave were also carried out which show 
new variety of nonlinear state in which zonal and streamer 
powers were comparable. It was also shown that the parameter regimes for which one obtains unsaturated 
streamer patterns could be stabilized by increasing the aspect ratio $L_x/L_y$ from unity. 
This basically enables the secondary KH destabilization of the streamer patterns. 
It was shown that such cases would be responsible for  higher transport.

\section{Simulation of the 3d electromagnetic model}
We now turn to the generalized three dimensional model represented by Eqs.(\ref{eq:1}-\ref{eq:3}) and discuss 
its numerical solutions. 
In comparison to the 3d electrostatic case we now have an additional 
field variable $\psi$ (the magnetic vector potential), whose   temporal evolution is governed 
 by Eq.(\ref{eq:3}) and which  
provides additional coupling  (linear and nonlinear both) terms in the evolution equations of the
density and potential variables. The coupling coefficient is proportional to $V_A^2$ 
(i.e. inversely proportional to $\beta$) as well as to the spatial variation in the parallel direction 
(i.e. to $k_{\parallel}$). As mentioned before the influence of 
electromagnetic effects will be felt 
 when $\partial \psi/\partial t $ becomes 
comparable to $\eta_s V_A^2 \gr^2 \psi$. The order of typical perpendicular wavenumbers 
ranging from $0.1$ to  unity for our simulations, it implies a 
direct comparison of $\eta_s V_A^2 $ with the growth rate. When $\eta_s V_A^2 >> \gamma $,    $\psi$ is essentially 
damped, electromagnetic efects are weak  and energy in the magnetic field is  typically small. This is the limit 
where the generalized 3d electromagnetic model is expected to reduce to the 3d electrostatic model discussed 
in detail in the last section. The parameter regime of $\eta_s V_A^2 << \gamma $ to $\eta_s V_A^2 $ 
of the order of $\gamma$ is 
thus of interest for  studying the influence of electomagnetic effects on the dynamics. 

In this section we present results for the  case of $\eta_s V_A^2 = 1$. For this value of $\eta_sV_A^2$ we expect the 
electromagnetic effects to become  significant enough so as to influence the dynamics. 
Furthermore, this choice is also motivated by the fact that it is the  parameter regime relevant  for 
the currentless toroidal BETA machine at IPR. A detailed parametric simulation study  for   $\eta_s V_A^2$ 
ranging from $ << \gamma $ to $ >> \gamma $
for the 3d electromagnetic set of equations  are, however, underway and will be presented  elsewhere.

 We look for saturated states in a square  box geometry with ($L_{x}=L_{y}$). The other parameters are 
 $V_g = 0.036, V_n = 0.8, D = \mu = 0.1$ and $2\pi V_A/L_z = 10^{-3} $. 
The plot in Fig.13 shows the evolution of the total and the magnetic  energy (solid and dashed lines respectively). 
The energy is seen to saturate 
after an initial exponential growth. The saturated magnetic energy is an order of magnitude smaller than the 
total energy for this set of parameter values. Thus the electromagnetic fluctuations are  not dominant 
in this particular smulation. However, we  present a comparison with the electrostatic case to show that 
even in  this case, though  the amplitude  of electromagnetic fluctuations are low, their effect on the 
nonlinear saturated state is significant.  The linear growth of energy for both electromagnetic and electrostatic cases 
are similar. This is because the value of the maximum growth rate in both the cases are identical. However, in 
the nonlinear regime the electrostatic total energy is smaller due to  a mild decay during the later phase.  
The lower subplot of the same figure shows the evolution of intensity  of  zonal 
and the streamer modes. The zonal intensity for both electrostatic and the 
electromagnetic cases are at  identical level. The intensity of the streamer mode, however, in the electrostatic 
case is considerably smaller and exhibits a mild decay similar to what is  observed for the total energy. 
It is interesting to  note  that in the  electrostatic case even though the total energy 
in turbulent fluctuations are small compared to the 
electromagnetic case, the zonal intensity is identical, {\it i.e.} as a relative 
fraction, zonal flow intensity is stronger in the electrostatic case. This clearly implies that it is much easier 
to generate zonal flows in the electrostatic case, confirming the prevalent lore that 
that electromagnetic effects  inhibit the zonal flow generation. The streamer intensity in electromagnetic case 
being higher also confirms  that the stabilization of the instability in the presence of electromagnetic 
effects becomes difficult.

In Fig.14 we depict  the slice plots for the three fields viz. $n$, $\vph$ and $\psi$
 in both the linear as well as the  nonlinear regimes. In both linear and nonlinear  regimes we  
observe  considerable structure in all the three fields in the $z$ direction, indicating that 
energy in the finite $k_z$ modes are 
significant.   Unlike the clear symmetry breaking nonlinear stages (growing streamers or saturated zonal 
depending on the parameter regimes in the $D, \mu $ and $\eta_s$ space) of the 
3d electrostatic RT modes, in the electromagnetic case the flow structures cannot be 
distinctly classified in  zonal and streamer 
patterns. This is also evident from the evolution of zonal and streamer intensities in the plot of Fig.13. 
Furthermore, the $n$ and $\psi$ fields 
show a predominance of short structures in comparison to the potential $\vph$ 
field. 

Our simulations also show that similar to 3d electrostatic case, here too
the density remains an independent field 
throughout the evolution and does not get slaved to the potential field. 
In the 3d electrostatic context, this was attributed to the 
the presence of a variety of additional modes arising by permitting the  three  dimensional variation 
in  the system   
and also to the loss of integral invariant for the 3d equations. In the electromagnetic case where 
there is an increase 
in the variety of linear modes, (shown  in detail in the third section ) it is even more difficult 
for the two fields to develop any functional relationships.  

The electromagnetic studies of this section  reveal that the features, which were 
earlier (with the help of electrostatic  studies), attributed to the three dimensionality of the system 
are present in these simulations also, the sytem being three dimensional here as well. 
However, additionally we observe that even the presence of a weak electromagnetic energy 
considerably opposes the process of nonlinear stabilization.

\section{Summary and conclusions}
In this work we have studied the extension  of a 
previous two dimensional nonlinear model evolution 
equations \cite{rtipr2,rtipr1} for the magnetic curvature driven Rayleigh 
Taylor instability to three dimensional perturbations. The extended model  also 
incorporates  coupling to electromagnetic fluctuations 
associated with the magnetic field line bending terms through the parallel component of the Ohm's law. 
The objective of the present work has been to investigate the influence of three 
dimensionality and the electromagnetic effects  on the nonlinear state. 
It was shown that the effects due to three dimensionality can be isolated by considering 
a simplified 3d electrostatic limit. Such a limit is  valid when the typical growth rates are much smaller than 
the parameter $\eta_s k_{\perp}^2 V_A^2$. Here $\eta_s$ is the resistivity parameter defined earlier in the text,  
$ \beta = 1/V_A^2$ represents the plasma beta and $k_{\perp}$, the typical perpendicular scales.  

Studies on 3d electrostatic model show that the  $D - \mu$ phase space, 
which in  earlier  2d studies \cite{rtipr2} governed the symmetry of  nonlinear flow patterns 
(viz. the transport inhibiting saturated  
zonal flows versus  the transport enhancing growing streamers) 
 gets extended by the inclusion of a third dissipative parameter viz.  $\eta_s$. 
A comprehensive parametric study reveals that the resistivity parameter, 
has an entirely different role in pattern selection process. While smaller values of both $D$ and $\mu$ 
form zonals and their larger values streamers, it is the opposite for $\eta_s$. 
In 2d simulations it was observed that the density field ultimately develops a 
functional relationship with potential. The power cascade towards long scale for the potential in 2d case 
ultimately also forces the density to acquire long scale structures. It was observed in our current 3d simulations 
that the density was not in any way constrained to follow the potential field. Hence, in the 3d case the density 
field continues to have power in short scale fluctuations. Another feature of the 3d model, in 
contrast 
to 2d,  is  the existence of  additional resistivity driven modes. Nonlinear studies on  resistively 
destabilized drift wave (with  RT growth  switched off) 
yielded a novel variety of saturated states which have neither zonal nor streamer symmetries; 
instead the intensities of both zonal and streamer flows 
were at  comparable level. 

The  electromagnetic effects on RT were studied by simulating the fully 3d electromagnetic set of 
equations. At the moment we have carried out investigations only for those parameters 
for which the electromagnetic effects are  
weak and  the magnetic energy is an order of magnitude smaller than the total energy. This was achieved by choosing 
$\eta_s V_A^2 = 1$, which is still larger than the 
maximum growth rate $\gamma_{max} = 0.16$. 
These simulations clearly show that even though the electromagnetic energy is weak, its presence 
inhibits the zonal formation leading to a relatively higher amplitude of streamers. Thus, as expected, the 
presence of electromagnetic effects hinders the saturation process. 

Curvature or gravity driven RT modes are important in many magnetized plasma problems. They were extensively 
studied experimentally in the toroidal currentless plasma machine BETA \cite{beta}. As stated in the text, the parameters 
of BETA were such ($V_n = 0.8$, $V_g = 0.036$, $L_z = 1.25\times 10^5 \rho_s$, $\eta_s \sim  10^{-5}$) 
that the results of present study are directly applicable. In the context of tokamaks, curvature driven 
ballooning modes (with or without resistivity effects ) are relevant to the core region, as well as the edge region 
(especially when the plasma ehibits the edge localized modes called ELMS). The typical parameters in the core 
region are ($\eta_s \sim 10^{-7}$, $V_A^2 = 1/\beta =10$, implying that $\eta_s V_A^2  \sim 10^{-6}$, 
$L_z/\rho_s = 2 \pi q R/\rho_s = 2 \times 10^4$, $V_n = \rho_s/L_n \approx 0.1$ and $V_g = \rho_s/R \approx 10^{-2}$)  
and those in the edge region are ($\eta_s \sim 2.5 \times 10^{-5}$, $V_A^2 = 1/\beta =10^3 - 10^4$, 
so  that $\eta_s V_A^2  \sim 1$, 
$L_z/\rho_s = 2 \pi q R/\rho_s = 2 \times 10^4$, $V_n = \rho_s/L_n \approx 0.1$ and $V_g = \rho_s/R \approx 10^{-3}$ ). 
Thus the present study may be applicable to the tokamak edge problem, 
although the driving instability in the simulations is somewhat stronger because of  the higher value 
chosen for $V_n$. Gravity driven RT modes are also relevant to the spread - F region of the ionosphere. 
The typical parameters in this region are ($\eta_s = 10^{-5}$, $V_A^2 = 1/\beta \approx 10^6$, $\eta_S V_A^2 \sim 10$, 
$L_z\rho_s = 10^5$, $v_n = \rho_s /L_n \approx 2.5 \times 10^{-3} $ and $V_g = 10^{-4}$). 
Thus the results of our studies are applicable to this problem also. 

It should be pointed out here that several three dimensional numerical studies on the electrostatic \cite{guz2,guz3} and 
electromagnetic \cite{dr1,dr2} nonlinear equations describing the coupling of drift ballooning modes have been carried out 
in the last decade or so. Such studies have employed  a realistic three dimensional model for tokamaks with effects due 
to magnetic shear and parallel flows. These studies contain extensive details of simulation studies on 
  transport  in tokamaks. Our objective in contrast has been to delineate the parameter regime for the formation 
of transport inhibiting and transport enhancing structures and identification of the rudimentary physics with the choice of a 
simplified model ( that of drift - Rayliegh Taylor coupling, in the absence of both parallel flow and magnetic shear).    

Finally, we make some remarks on the further exploration of the present  work   that we are currently  pursuing. 
The complete parametric study for the fully 3d electromagnetic case has not been presented here. 
Future studies will explore other regions of parameter space (especially lower values of $\eta_s V_A^2$ 
where electromagnetic effects become more important). Thus   
future investigations will be carried out 
to understand the  nonlinear stabilization process by the zonal flow formation as the  
the parameters $\eta_s V_A^2$ and $\beta = 1/V_A^2$ are varied. We also want to include the finite $T_i$ effects, 
bacuse finite ion Larmor radius stabilization is an important linear mechanism of the stabilization of the curvature 
driven instabilities.  

\vspace{0.2in}
\noindent
{\bf Acknowledgement:} We are thankful to Xavier Garbet and other 
organizers of the workshop on `` Relaxations in magnetized plasmas, 7 - 25 July 2003'' 
held at Aix - en - Provence, France, where part of this work was carried out.

\newpage
\begin{center}
FIGURE CAPTIONS
\end{center}
\vspace{0.2 in}
\begin{itemize}
\item[Fig.1]
Plot of the real frequency ($\omega_r$) and the growth rate $\omega_i$ vs. $k_zV_A$ 
for the electromagnetic dispersion relation of Eq.(\ref{dispersion}) in the non-dissipative limit i.e. 
for $D = \mu = \eta_s = 0$. The other parameters  are $V_g = 0.036$, $V_n = 0.8$.  
The perpendicular scales are $k_x = k_y = 0.1$. 
\item[Fig.2]
Plot of real frequency and the growth rate from the dispersion relation of Eq.(\ref{dispersion}) as a function of $k_zV_A$ 
when $\mu = 3$ is finite (subplot(a) and (b)) and when $\eta_s V_A^2 = 1.6$ is finite  (subplot(c) and (d)). The  
other parameters are same as that of Fig.1. 
\item[Fig.3]
The plot of real (solid lines) and imaginary (dashed lines)  part of the 
frequency for the resistivity driven  branch.  The ideal results for $\eta_s  = 0$ are plotted in subplot(a) ($V_g = 0.0364$) 
and subplot(c) ($V_g = 0.01$) for the purpose of comparison  with the finite resistivity $\eta_s V_A^2 = 1.6$ plots of subplots (b) 
and (d) respectively. The other parameters are the same as that of Fig.1. 
\item[Fig.4]
The two subplots show   growth rate vs. $k_zV_A$ for resistive destabilized drift wave $V_g = 0$; (the other parameters being 
$V_n = 0.8$, , $k_x = k_y = 0.1$, $D = \mu = 0$) from the fully 3d electromagnetic dispersion relation of 
Eq.(\ref{dispersion}) (solid lines, here ) and for the simplified 3d electrostatic dispersion 
relation of Eq.(\ref{es_disp})
(circles).  The growth rates for the two models differ at small $k_zV_A$ in the upper subplot for which 
resistivity parameter $\eta_sV_A^2 = 1.6$. They yield almost identical growth rates for large $\eta_sV_A^2 = 16$ as shown in 
the lower subplot. 
\item[Fig.5]
The slice plots (described in text) for potential $\vph$ field to visualize its three dimensional structure 
at various times for the 3d electrostatic evolution equations (\ref{eq_es:1}, \ref{eq_es:2}).   
The parameters for this case are $V_n = 0.8$, $V_g = 0.036$, $D = \mu = 0.1$. The simulation box sizes are 
$L_x = L_y = 20\pi \rho_s, L_z = 1.25 \times 10^5 \rho_s$
The first three plots (as indicated on the top) are for $\eta_s = 10^{-5}$ at $t = 0, 30 and 150$ and show the formation of zonal 
$y$ symmetric pattern in the final nonlinear state at $t = 150$. 
The fourth plot corresponds to a different run for which $\eta_s$ is   $ = 1.26 \times 10^{-6}$, and shows  
formation of radially (along $x$) extended streamer structure. 
\item[Fig.6]
The parametric regime of $D$ vs. $1/\eta_s$ space in which the circles indicate those values of $D$ and $1/\eta_s$ 
for which nonlinear saturation was achieved. The evolution is governed by  the 3d electrostatic set of equations 
(\ref{eq_es:1}, \ref{eq_es:2}). The other parameters have the same values as that chosen in Fig.5. 
\item[Fig.7]
Plot of the ratio $\gamma \eta_s/k_z^2$ vs. the permissible range of $k_z$ in  simulation for  
$ \eta_s= 1.26 \times 10^{-6}$ (dotted 
line) and  $ \eta_s= 10^{-5}$ (+ sign). 
\item[Fig.8]
The slice plots for density and vorticity $\gr^2 \phi$ in the nonlinear state for two different values of 
$\eta_s$ (viz. $10^{-5}$ and $1.26 \times 10^{-6}$).  The other parameters and the governing evolution 
equation are same as that of  Fig.5. 
\item[Fig.9]
Scatter plots between $n$ and $\vph$ and between $\gr^2 \vph$ and $\vph$ initially at $t = 0$ and at the stage 
where saturation is achieved at $t = 150$. The data for $n$ and $\vph$ is the same as that of Fig.5 with $\eta_s = 10^{-5}$. 
Clearly, the plots show no development of any kind of functional relationship unlike the 2d case \cite{rtipr1}.
\item[Fig.10]
The   numerical   evolution of  energy, streamer and zonal intensity per unit volume with time for the 
3d electrostatic model of Eq.(\ref{eq_es:1},\ref{eq_es:2}); for the Rayleigh Taylor (subplots in left column) 
and the resistive drift mode (subplots in the right column) for comparison.  The plots in solid lines show the intensity  
in $k_z =0$ modes and the dashed lines indicate the power in finite $k_z$ modes.  
\item[Fig.11]
Slice plots showing the three dimensional density and potential structures in the linear ($ t = 150$) and the 
nonlinear ($t = 450$) regimes for the resistively destabilized drift waves.  
\item[Fig.12]
The evolution of total energy/volume for the case when $L_x/L_y = 1$ (solid line) and when $L_x/L_y = 4$. 
\item[Fig.13]
The upper subplot shows the evolution of kinetic (solid line), pressure (dashed line) and the magnetic 
(dotted line) energy for the fully generalized electromagnetic simulations of Eqs.(\ref{eq:1} - \ref{eq:3}). 
The lower subplot shows evolution of power in zonal (solid line) and streamer modes (dashed line). 
\item[Fig.14]
The slice plots showing three dimensional patterns for $n$, $\phi$ and $\psi$ 
in the linear $t = 30$ and the nonlinear $t = 150$ regimes for the electromagnetic simulation 
corresponding to the saturated state ($\eta_s V_A^2 = 1$) of Fig.13.  

\end{itemize}

\newpage
\bfig
\centerline{\scalebox{1.0}{\includegraphics{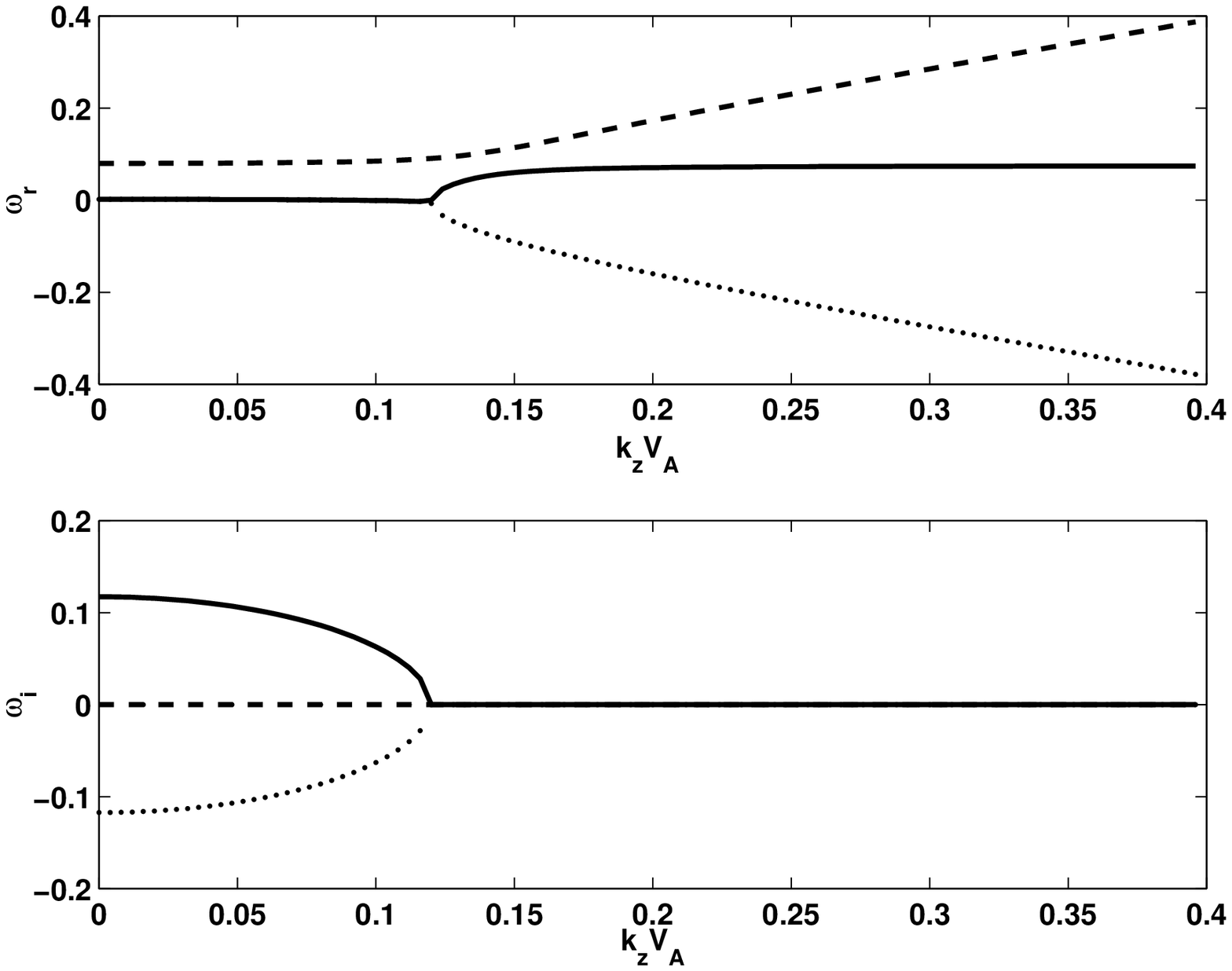}}} 
\caption{}
\efig

\newpage
\bfig
\centerline{\scalebox{1.0}{\includegraphics{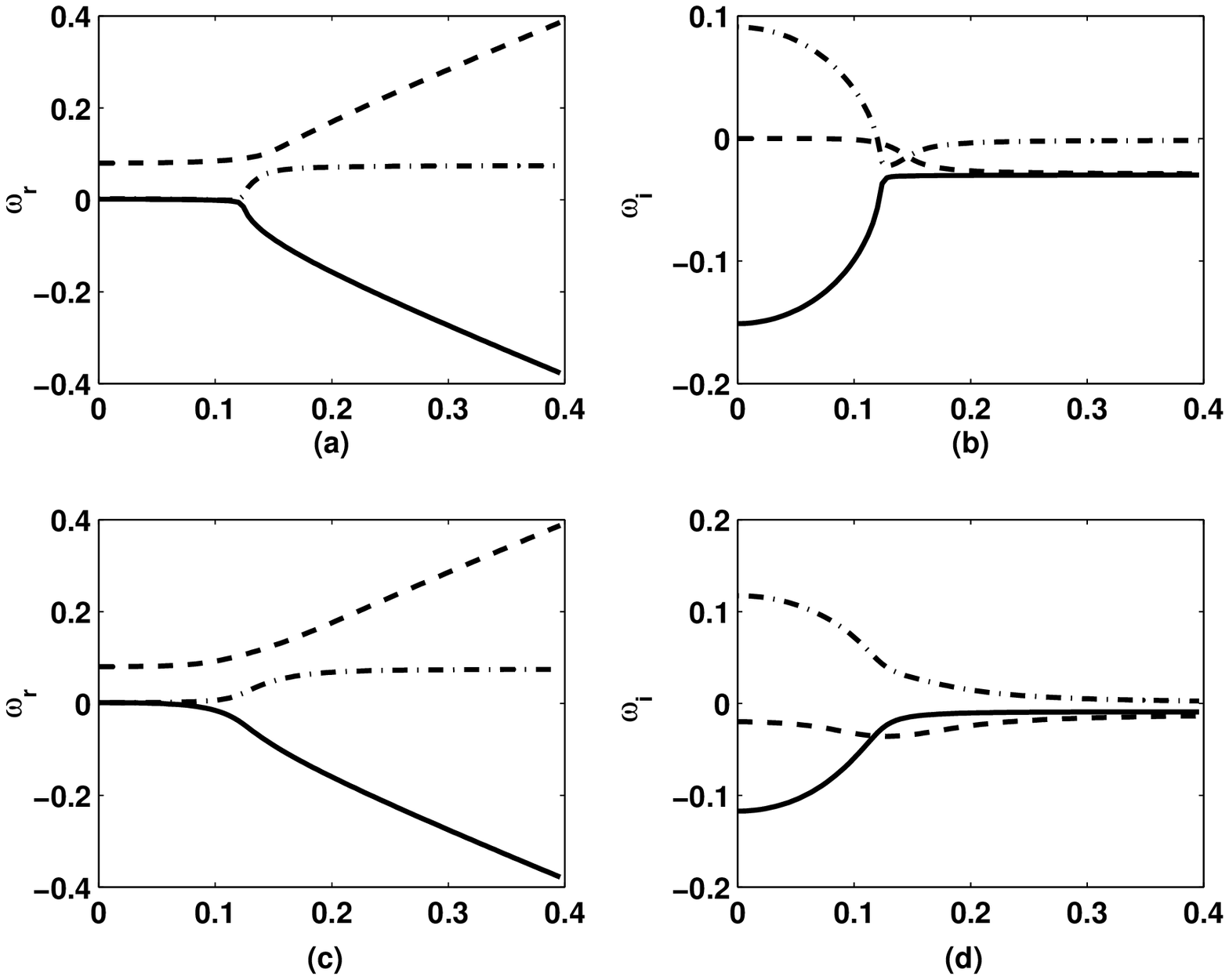}}} 
\caption{}
\efig

\newpage
\bfig
\centerline{\scalebox{1.0}{\includegraphics{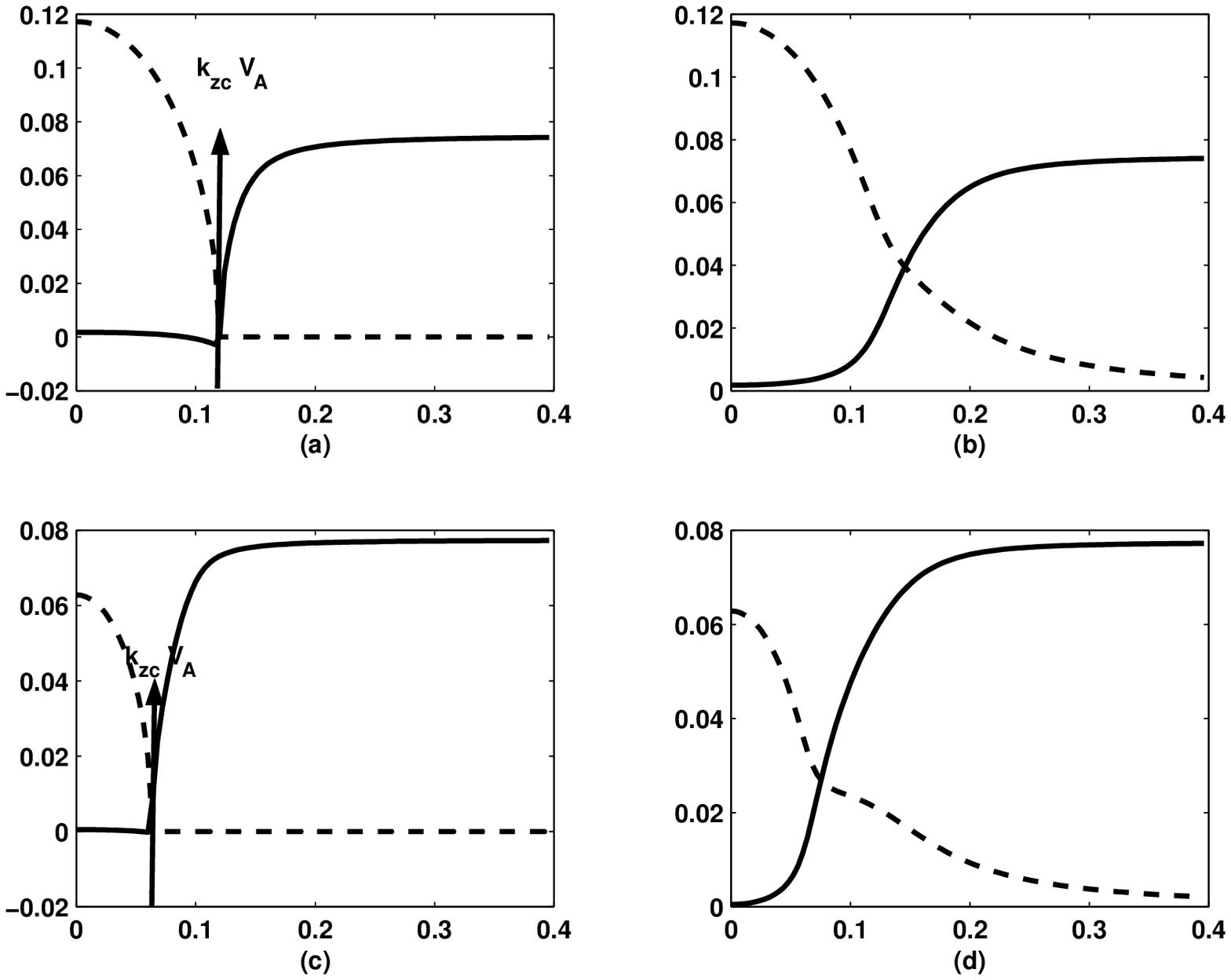}}}
\caption{}
\efig

\newpage
\bfig
\centerline{\scalebox{1.0}{\includegraphics{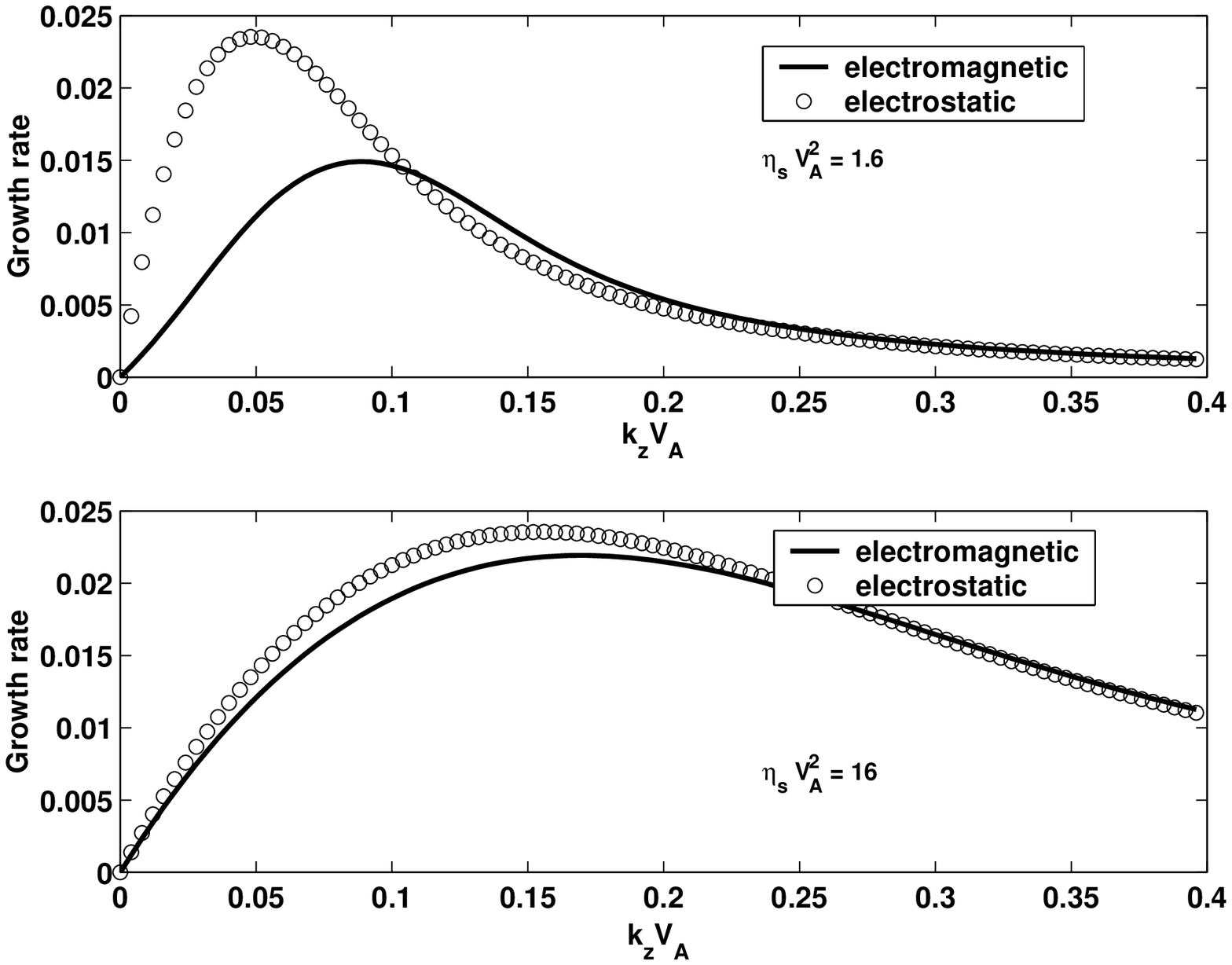}}}
\caption{}
\efig

\newpage
\bfig
\centerline{\scalebox{1.0}{\includegraphics{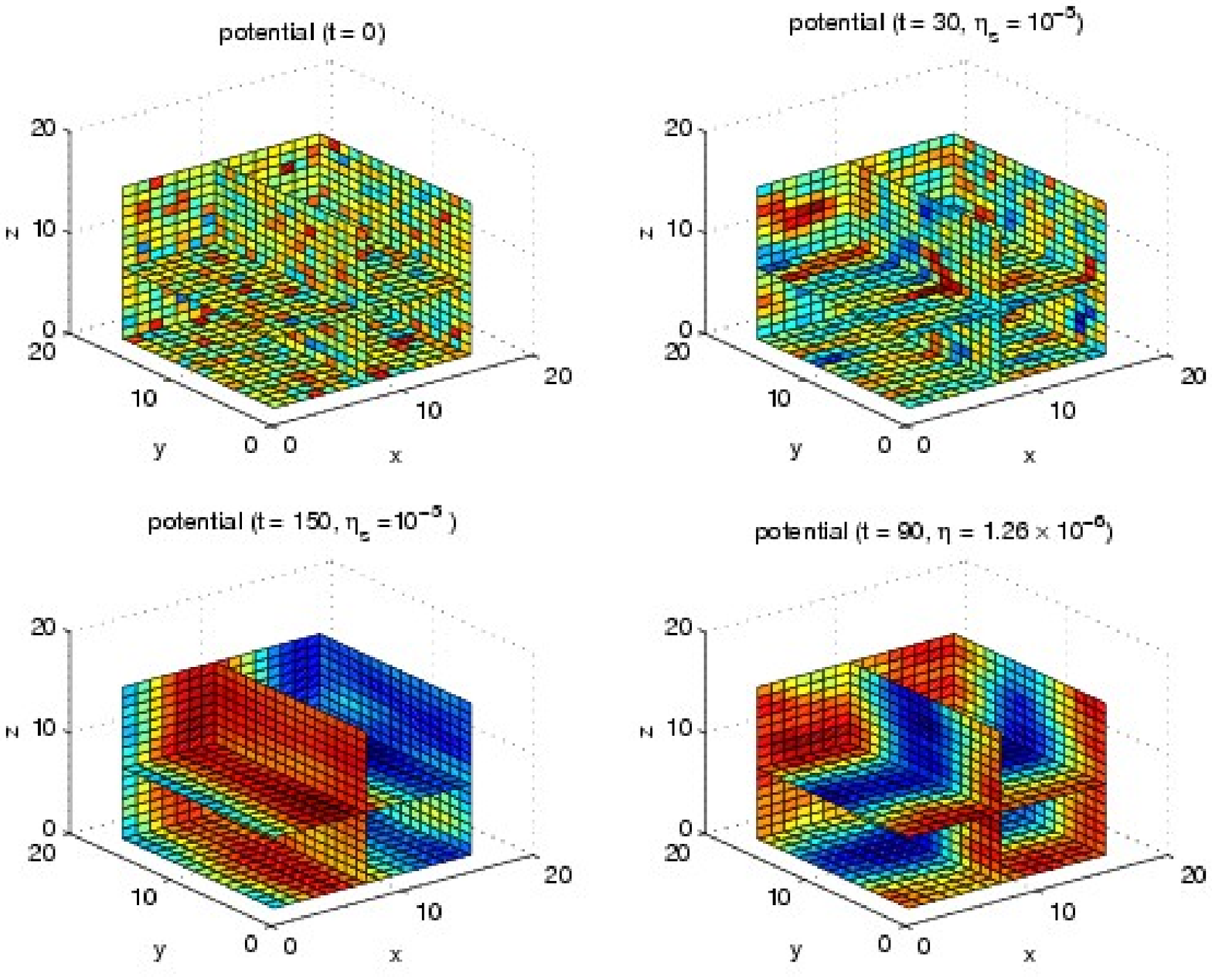}}}
\caption{}
\efig

\newpage
\bfig
\centerline{\scalebox{1.0}{\includegraphics{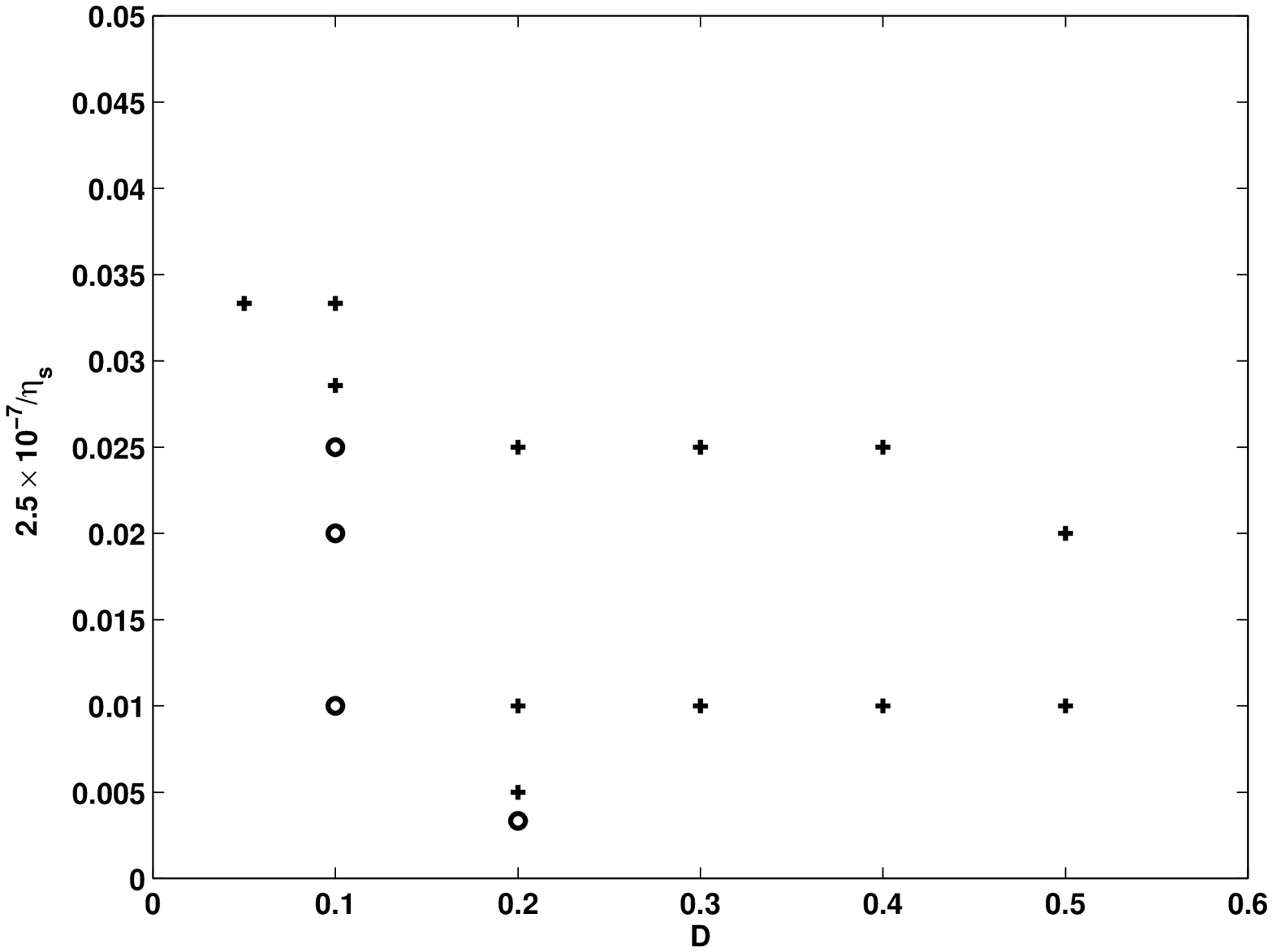}}}
\caption{}
\efig

\newpage
\bfig
\centerline{\scalebox{1.0}{\includegraphics{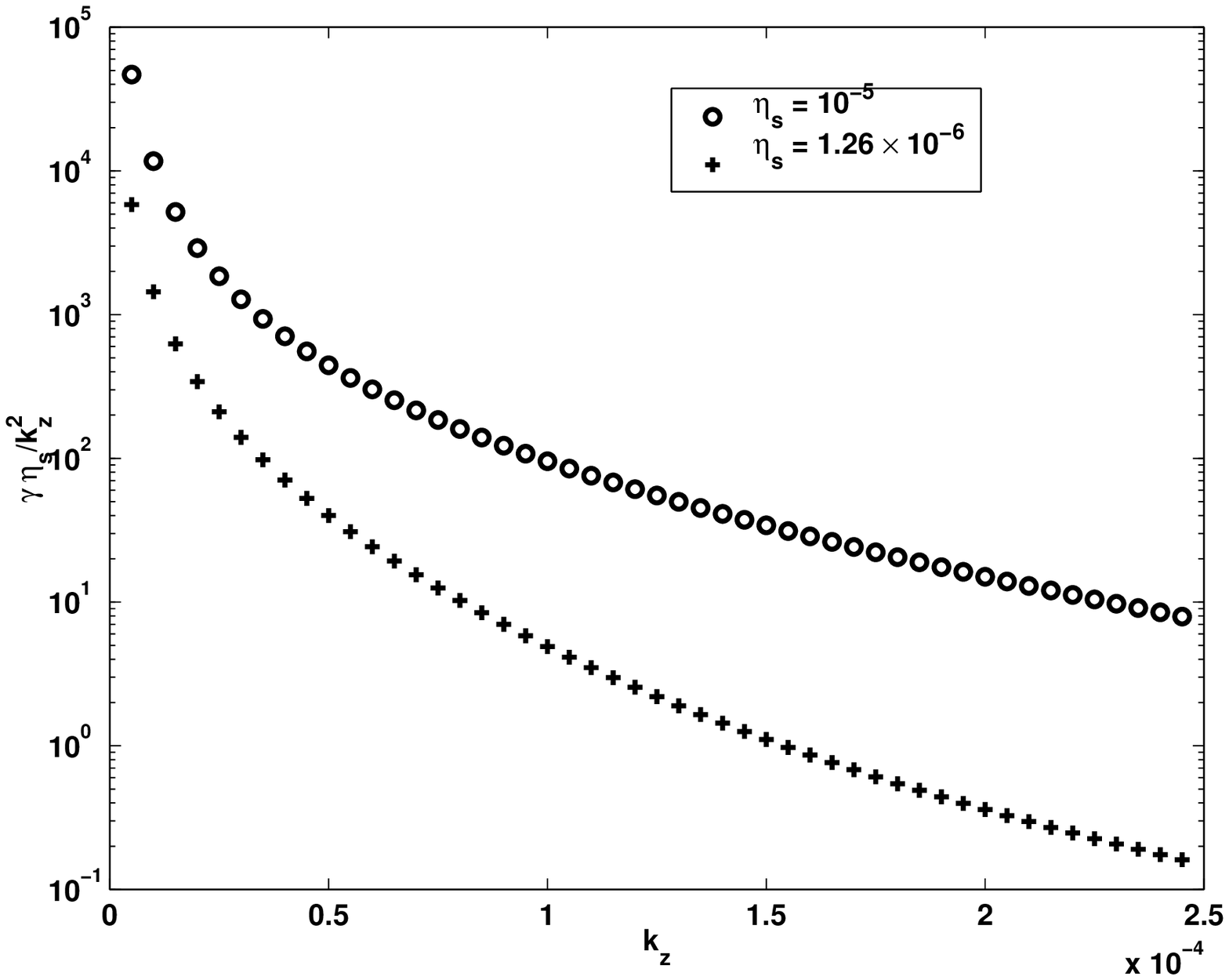}}}
\caption{}
\efig

\newpage
\bfig
\centerline{\scalebox{1.0}{\includegraphics{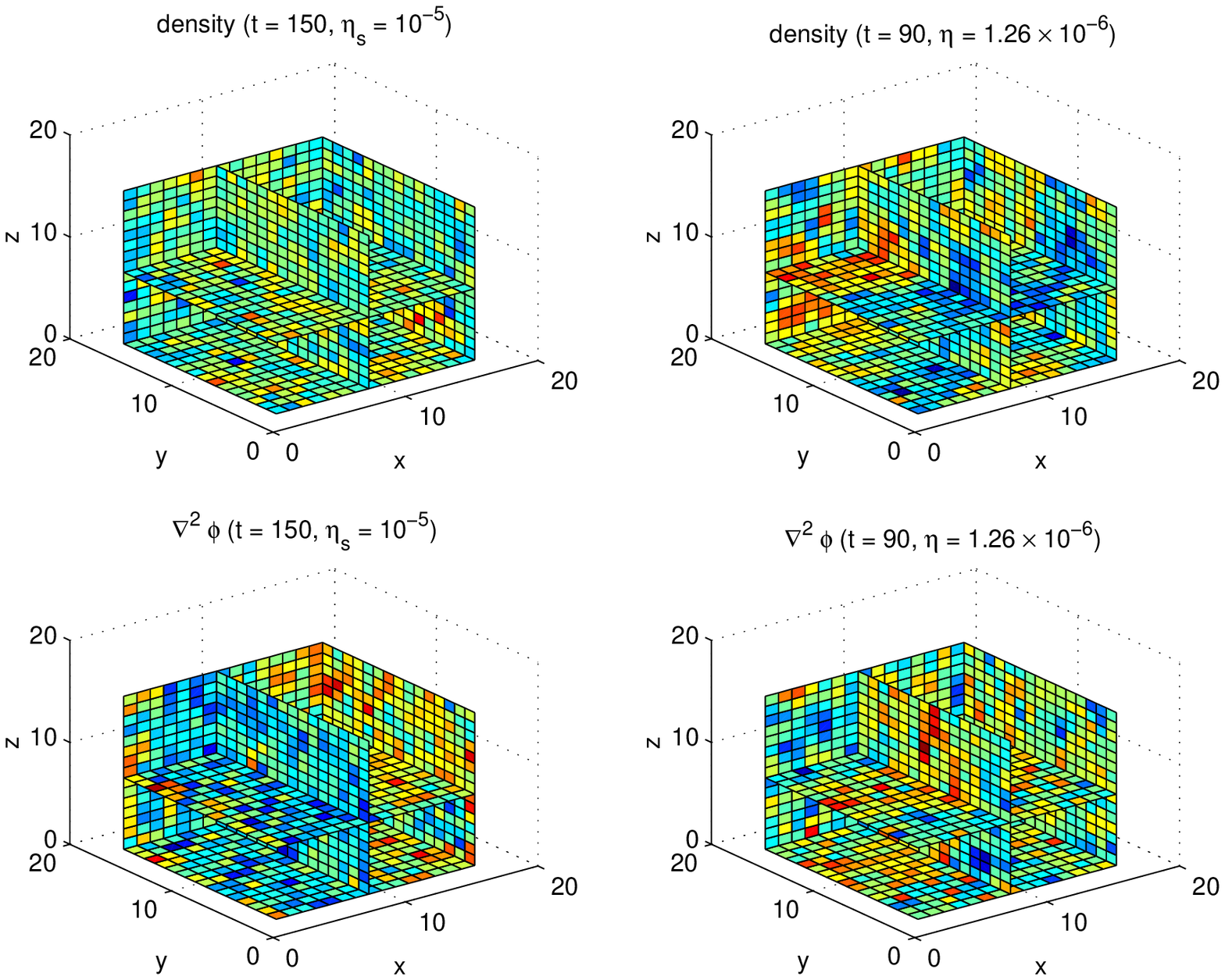}}}
\caption{}
\efig

\newpage
\bfig
\centerline{\scalebox{1.0}{\includegraphics{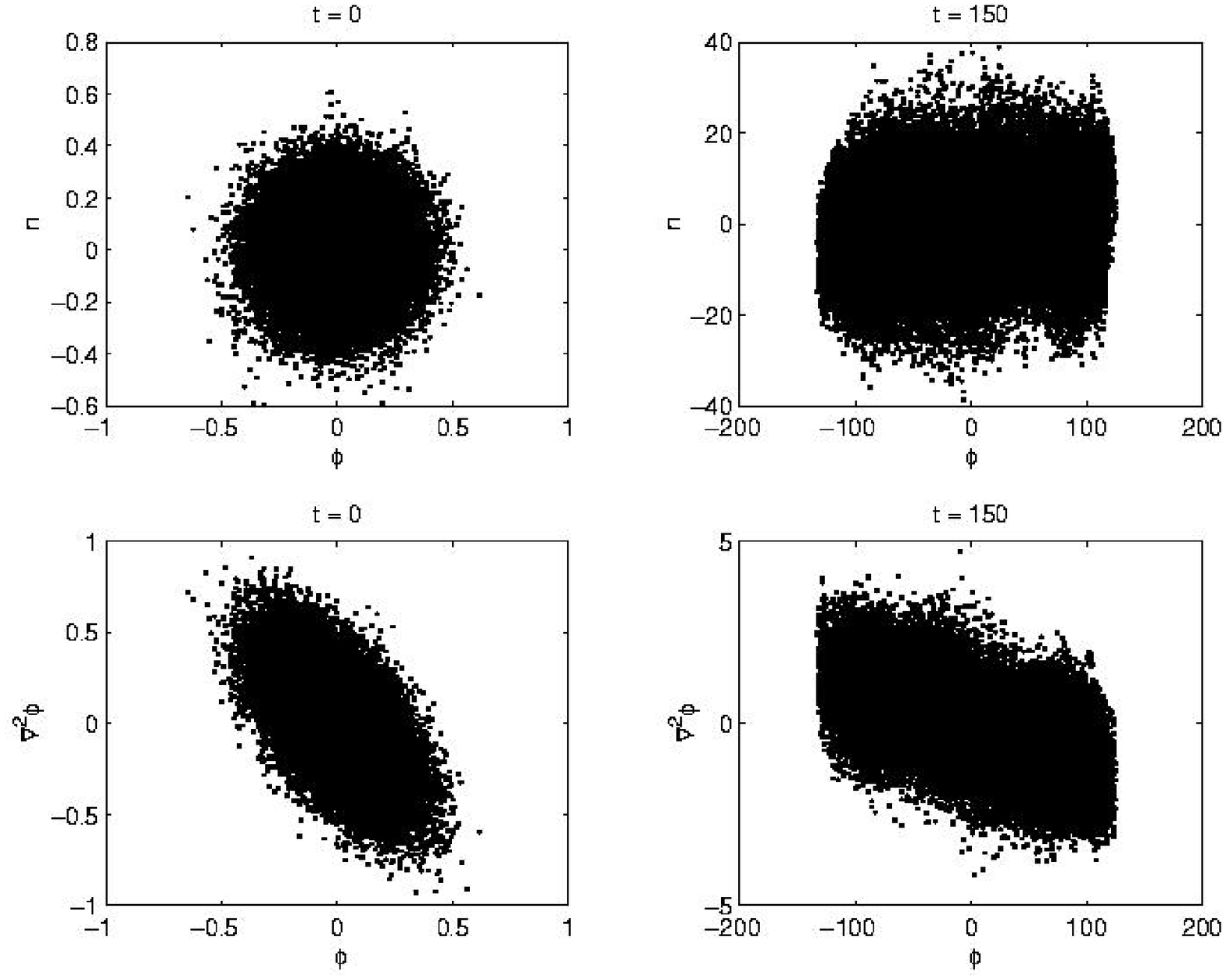}}}
\caption{}
\efig

\newpage
\bfig
\centerline{\scalebox{1.0}{\includegraphics{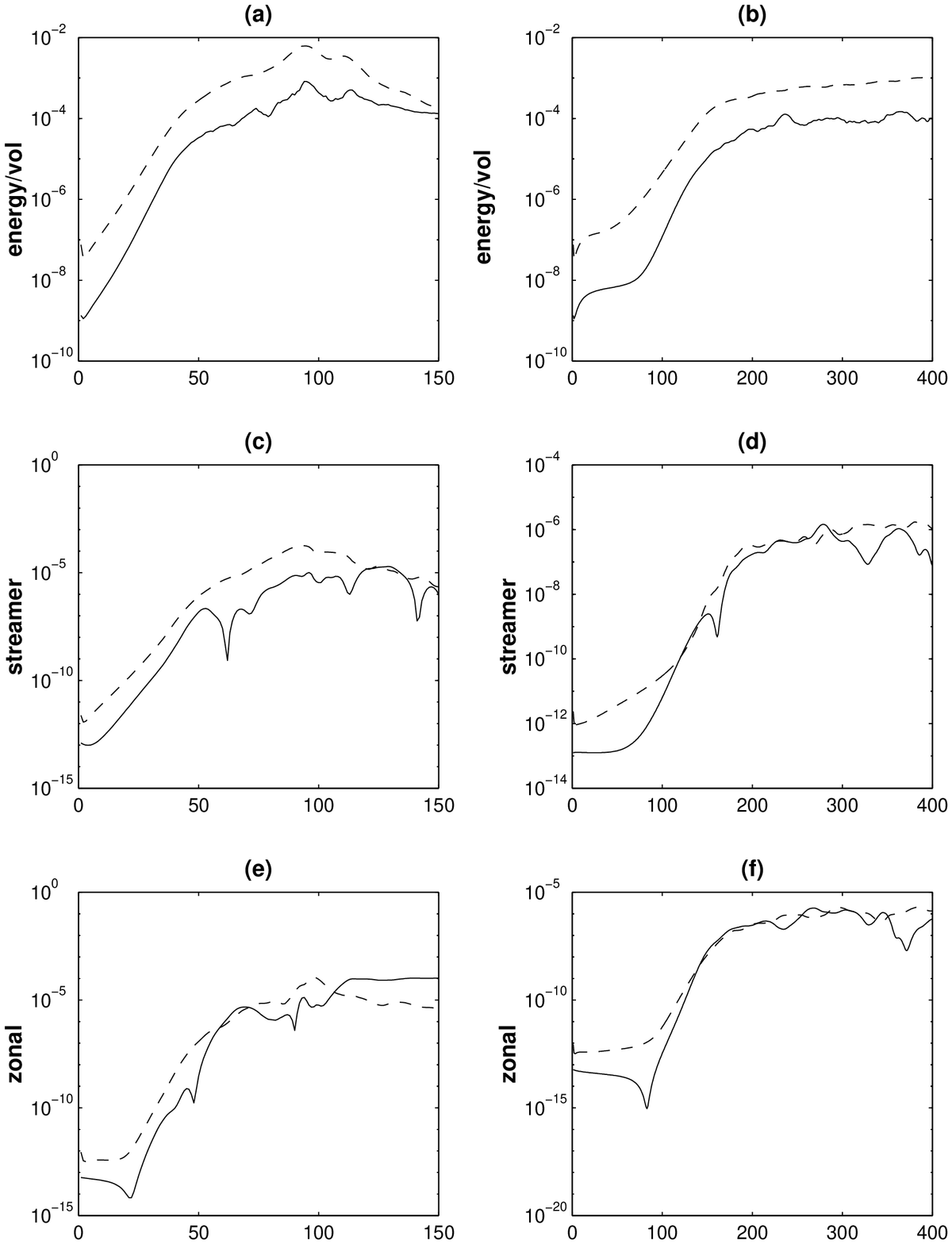}}}
\caption{}
\efig

\newpage
\bfig
\centerline{\scalebox{1.0}{\includegraphics{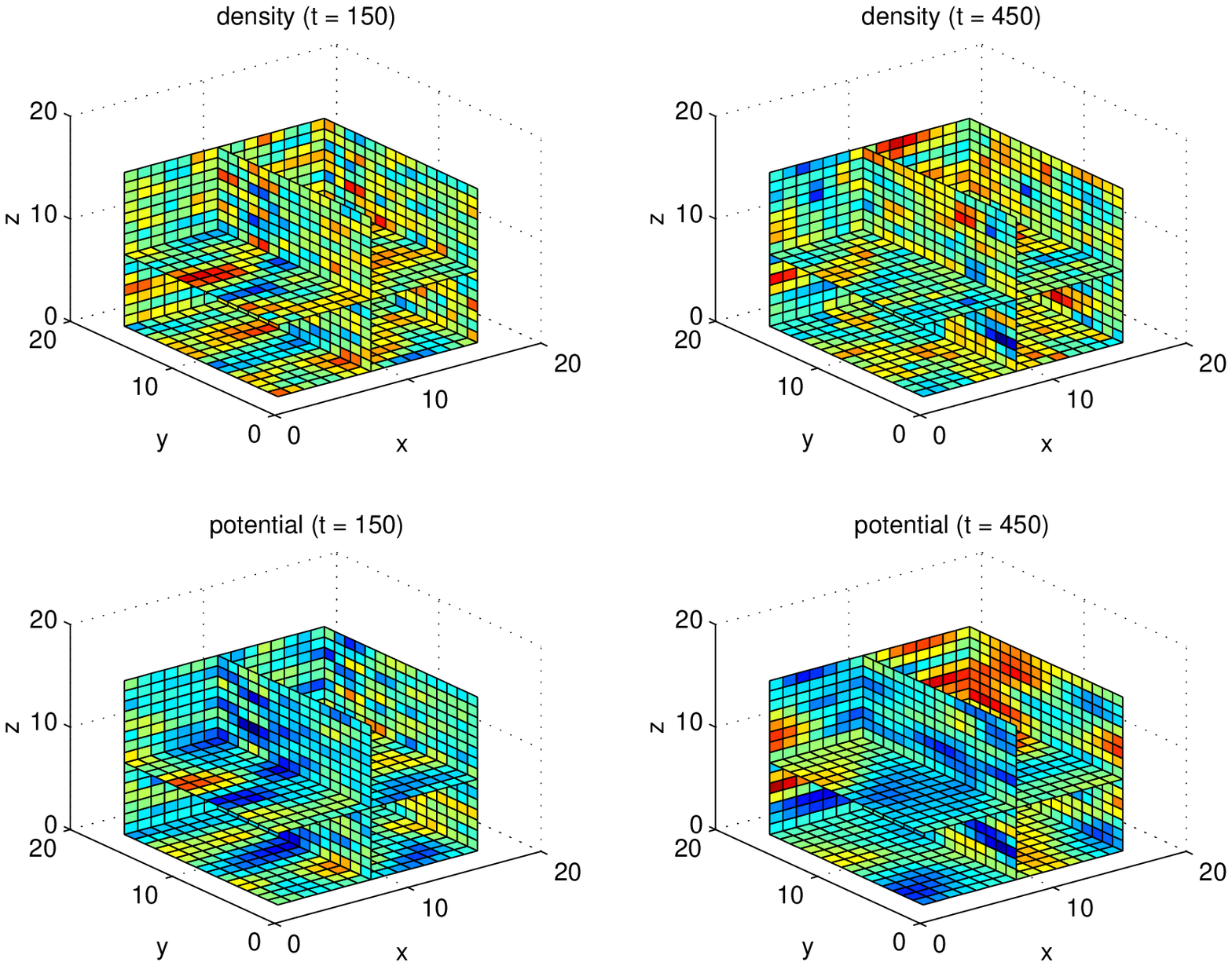}}}
\caption{}
\efig

\newpage
\bfig
\centerline{\scalebox{1.0}{\includegraphics{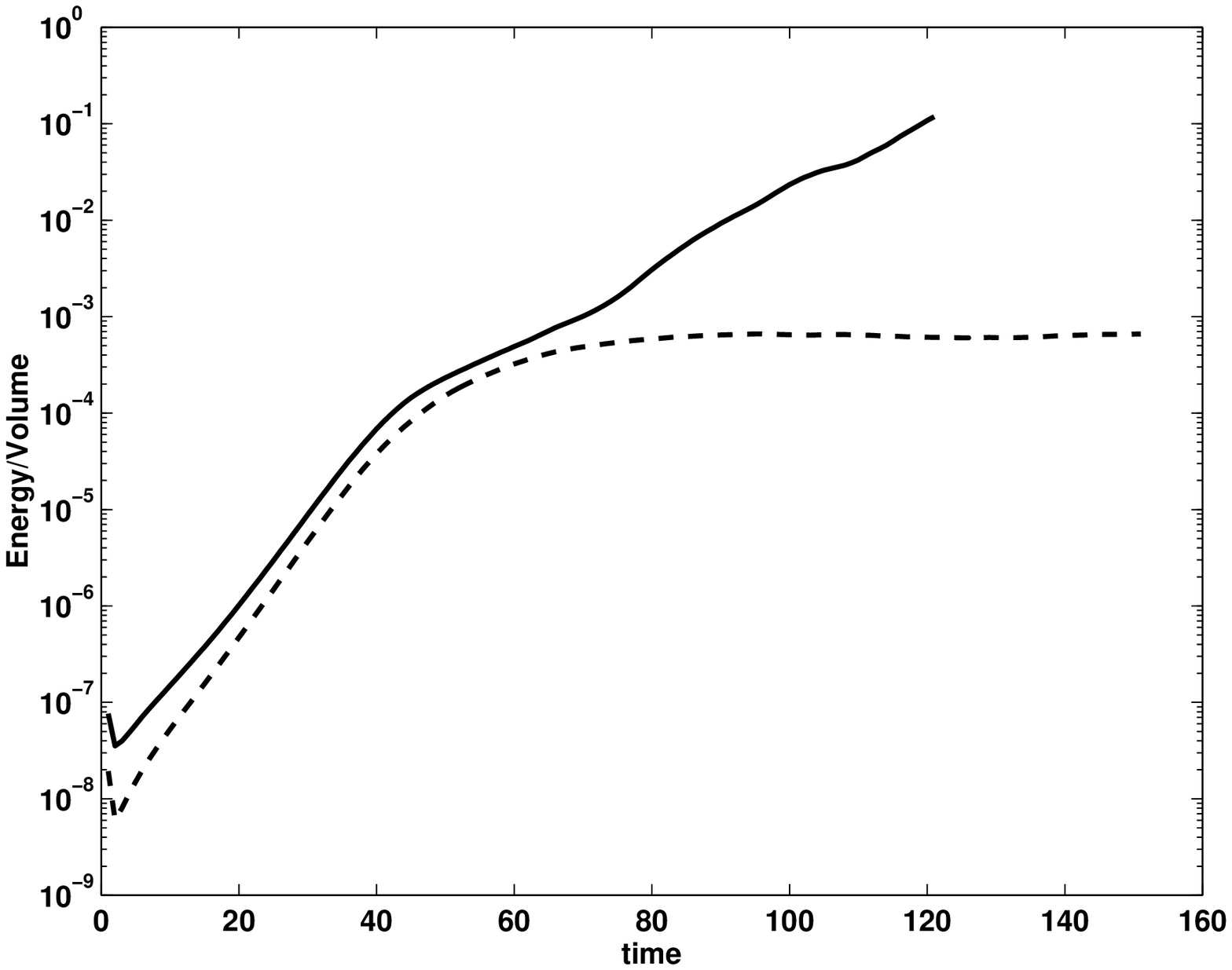}}}
\caption{}
\efig

\newpage
\bfig
\centerline{\scalebox{1.0}{\includegraphics{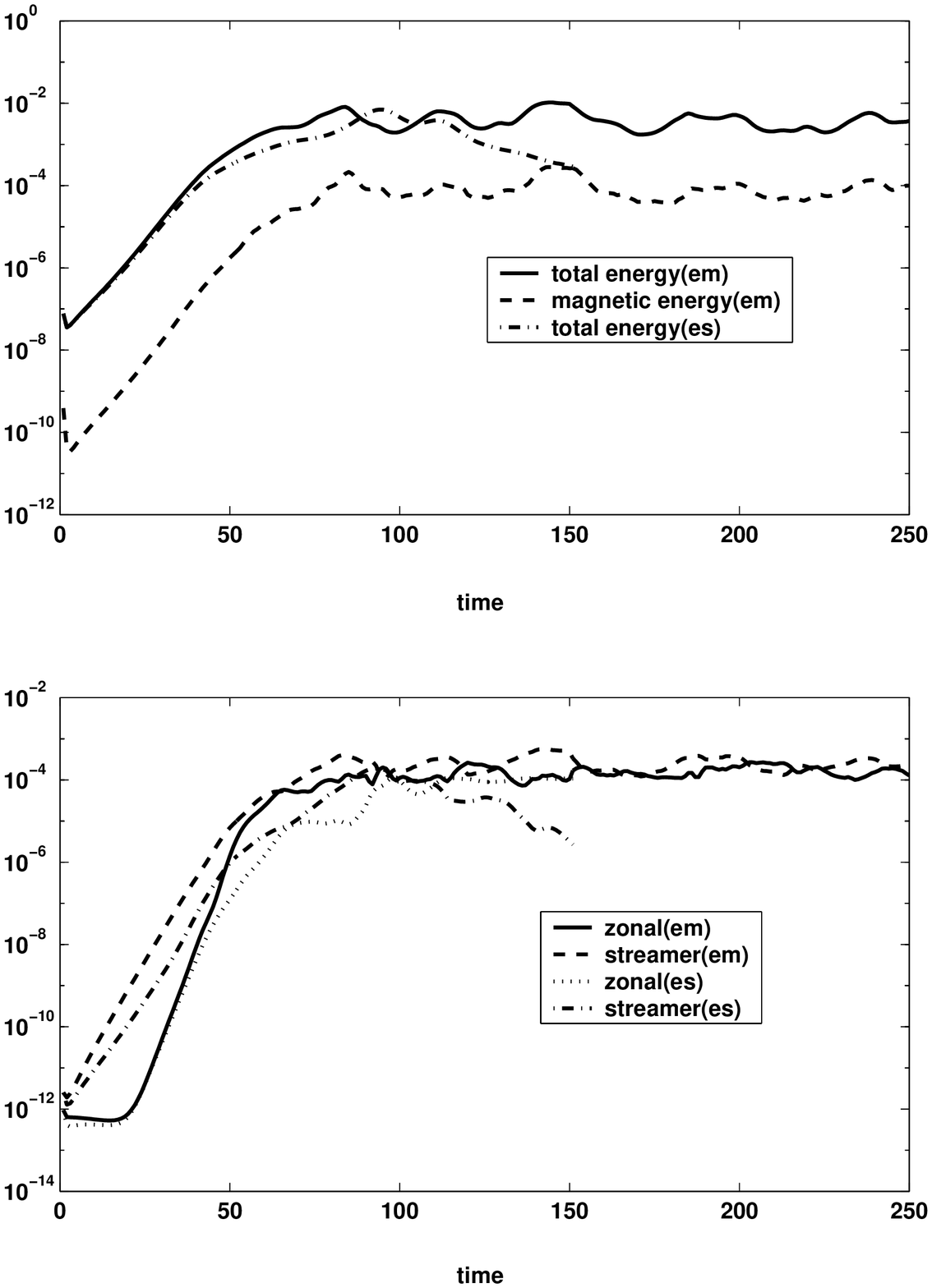}}}
\caption{}
\efig

\newpage
\bfig
\centerline{\scalebox{1.0}{\includegraphics{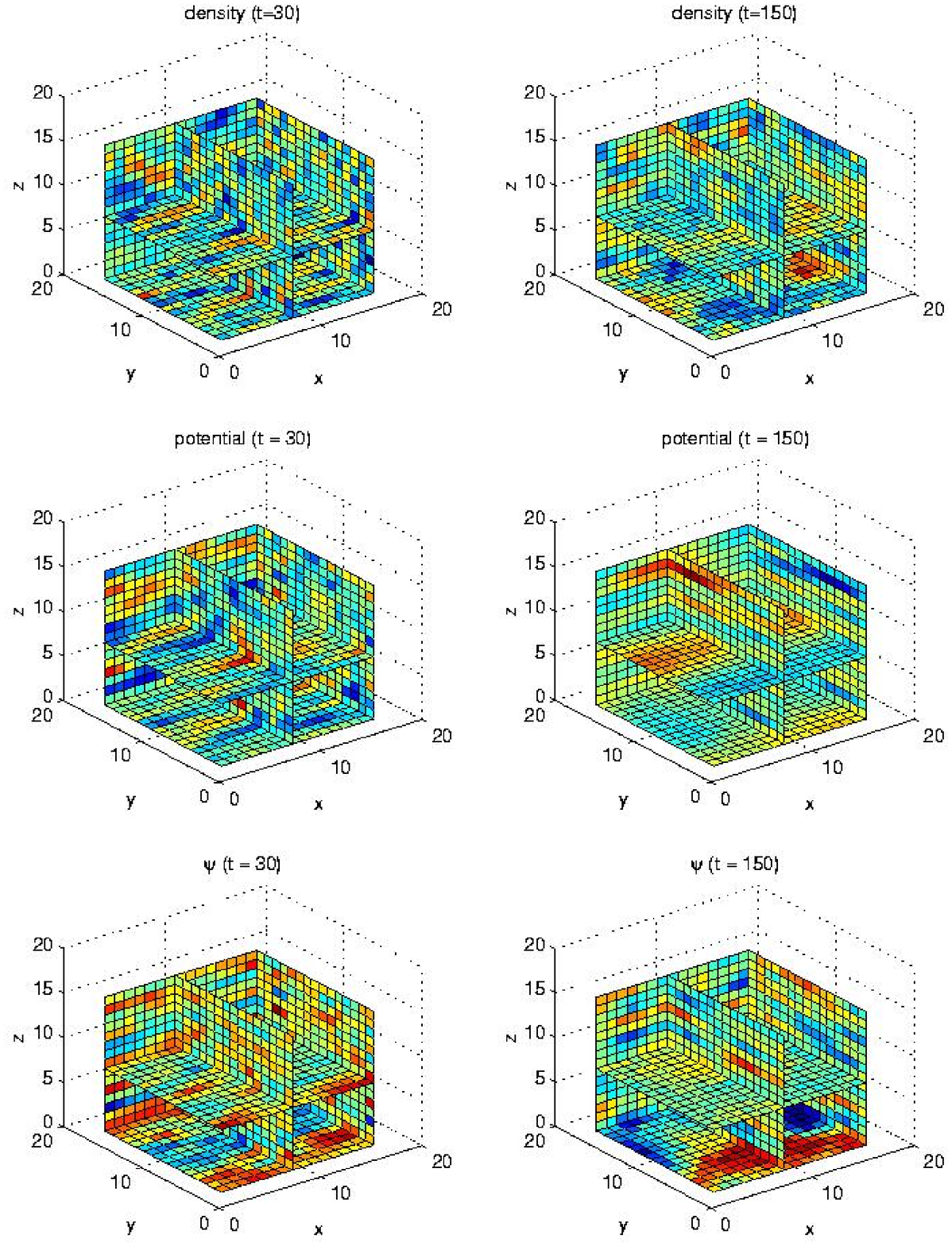}}}
\caption{}
\efig

\end{document}